\pdfoutput=1
\documentclass[11pt,a4paper]{article}
\usepackage[a4paper,margin=2.5cm]{geometry}
\usepackage[T1]{fontenc}
\usepackage[utf8]{inputenc}
\usepackage[british]{babel}
\usepackage{newtxtext,newtxmath}
\usepackage{microtype}
\usepackage{setspace}
\usepackage{graphicx,booktabs,longtable,array,float,multirow,tabularx,makecell}
\usepackage{caption,subcaption}
\usepackage[table]{xcolor}
\usepackage{tikz}
\usetikzlibrary{arrows.meta,positioning,shapes.geometric,fit,calc}
\usepackage{amsmath,bm}
\usepackage{enumitem}
\setlist{nosep,leftmargin=1.8em}
\usepackage[numbers,sort&compress]{natbib}
\usepackage{xurl}
\usepackage[hidelinks,pdfencoding=auto]{hyperref}

\title{Large Language Models for Forecasting of Electricity Price Spikes}
\author{Liang Lin\textsuperscript{1}, Jianguo Zhu\textsuperscript{1}, Gang Lei\textsuperscript{2}, Allen Wang\textsuperscript{3}, Xin Lu\textsuperscript{4}\\[0.5em]
\small\textsuperscript{1}School of Electrical and Computer Engineering, The University of Sydney, Australia\\
\small\textsuperscript{2}School of Electrical and Data Engineering, University of Technology Sydney, Australia\\
\small\textsuperscript{3}Wood Mackenzie\\
\small\textsuperscript{4}CSIRO, Australia}
\date{}

\begin{document}
\maketitle
\begin{abstract}
Can market notices read by a large language model (LLM) improve short-horizon electricity price spike forecasts? We study five regions of Australia's National Electricity Market. A local LLM extracted features once per notice, only notices visible at the forecast origin were used, and a gradient-boosting model produced onset probabilities. In an out-of-sample development back-test (July 2022 to July 2024), notice features did not lower the 60-minute Brier score of a numeric model including the market operator's next-hour forecasts. At 24 hours they did, but rule-based features did so almost equally and attribution to notice content was not established. A locally frozen protocol, not publicly deposited, governed a retrospective single-use holdout from 15 August to 17 October 2026. At 60 minutes, the hypothesis that LLM features lower the numeric model's Brier score was not established (H1; relative Brier skill $-$0.09\%), so the planned comparison with rule-based features was not tested (H2). At 24 hours, the hypothesis that LLM features lower the score against a text-free control was not established (H3; $+$0.37\%, 95\% CI [$-$9.74, 1.39], inconclusive). By the frozen wording rule the non-significant 24-hour improvement comes mainly from recalibration and cannot be attributed to notice content, although post hoc the decomposition could not separate recalibration from resolution. The 24-hour base rate was 1.56\% on the holdout against 14.0\% in development. With 5 onsets on 3 days and 24-hour positives in 2 of 22 three-day blocks, the holdout neither established nor excluded a 24-hour effect.
\end{abstract}

\section{Introduction}\label{sec:intro}

Australia's National Electricity Market (NEM) interconnects five regions. Each region has its own spot price, updated every five minutes \citep{aemoAboutNEM2026,aemcSpotContractMarkets2026}, and settlement has also been at five-minute intervals since 1 October 2021 \citep{aemcFiveMinuteSettlement2021}. The price occasionally rises within minutes towards the market price cap, which is set administratively and reviewed by rule \citep{aemcMarketPriceSettings2023}. Such spikes are short, rare and hard to anticipate \citep{weronElectricityPriceForecasting2014}, and they tend to cluster when the power system is under stress \citep{christensenItNeverRains2009}. Retailers and contracted generators are exposed to them \citep{aemcSpotContractMarkets2026}, and these participants need a forecast of risk: how likely is a spike to begin within the next hour or the next day?

The Australian Energy Market Operator (AEMO) publishes short-horizon price forecasts, five-minute pre-dispatch (P5MIN) and pre-dispatch (PD) \citep{aemoMMSP5MIN,aemoPredispatchProcedure2026}, and market notices \citep{aemoMMSMarketNotice}. In the data used here these notices report reserve shortfalls, outages, constraint changes and interventions in text. A large language model (LLM) can read such notices without hand-written rules. We ask whether LLM-extracted notice features improve probability forecasts of spike onset beyond a numeric model that already uses AEMO's next-hour forecasts.

The studies reviewed in Section~\ref{sec:related} leave this question open. Removing the language model did not reduce the accuracy of three LLM-based forecasters \citep{tanLanguageModelsActually2024}. Among the studies reviewed here that add LLM-processed text to price forecasts, text is aligned by day or by a fixed lag, or its timing is not stated, and none reports a comparison fixed before the test data. The NEM spike models reviewed here use numeric inputs only.

We compare five specifications pooled over the five regions. A contains numeric features only, including P5MIN and PD summaries. A0 adds to A an increment model with no text features. B adds notice features produced by fixed rules. L1 adds the same kinds of features produced by an LLM. L2 adds to L1 an embedding of each notice. The evidence comes from a development back-test from July 2022 to July 2024 and a retrospective single-use holdout of forecast origins from 15 August to 17 October 2026.

We find that, over 365,622 development origins, notice features did not lower the 60-minute Brier score, and that the 24-hour reduction was matched almost equally by simple rules; the holdout was too small to settle the question. The statistical components are established. Relative to the studies reviewed in Section~\ref{sec:related}, what we add is their application to NEM spike targets and market notices within one design. (i) The target is the probability of spike onset within 60 minutes or 24 hours, scored by the Brier score, a strictly proper rule \citep{gneitingStrictlyProperScoring2007}, and notice visibility rests on a measured publication delay. A rule fixed in advance turned the delays of 43 notices (99th percentile 24.52 minutes) into a visibility delay of 25 minutes (Equation~\eqref{eq:visible}). This is a proxy, and the 24-hour hypothesis has no sensitivity analysis for it. (ii) Two controls address attribution. The rule-based baseline B allows LLM extraction to be compared with fixed rules applied to the same notices. The text-free control A0 is meant to separate text information from recalibration; because the final 24-hour A0 stopped at one tree and L1 at 8 to 16, it does so only imperfectly. (iii) The holdout was opened once under a frozen protocol. The protocol was frozen and hashed locally before opening and was not deposited in a public registry, so its timing carries no third-party timestamp.

The holdout had little power. It contained 5 onsets on 3 days, and the 24-hour positives fell in 2 of 22 three-day blocks. Of the three hypotheses (Table~\ref{tab:hypotheses}), H1, L1 against A at 60 minutes, was not established (relative Brier skill $-$0.09\%); the final 60-minute text increments are single trees, so H1 compares nearly identical forecasts. H2, L1 against B at 60 minutes, was not tested. H3, L1 against A0 at 24 hours, was not established ($+$0.37\%, 95\% CI [$-$9.74, 1.39]; inconclusive). The holdout therefore neither established nor excluded a 24-hour effect. By the frozen wording rule, for the non-significant point estimate the improvement comes mainly from recalibration and cannot be attributed to notice content, although post hoc the decomposition could not separate recalibration from resolution. H3 was added after a favourable development signal had been seen, and many of the 44 protocol amendments were prompted by development results or by the review of the opening procedure. Sections~\ref{sec:evaluation} and~\ref{sec:discussion} and Appendix~\ref{app:deviations} set out these and other disclosures.

\section{Related work}\label{sec:related}

This review is selective. It follows no systematic search protocol, and each statement about a gap refers only to the studies reviewed here. Authors of the present paper are also authors of two of the text studies below \citep{huangLLMEnhancedShortTermElectricity2025,luLargeLanguageModelBased2025}.

\subsection{Spike forecasting in the NEM}

Work on the NEM has modelled spike occurrence directly. \citet{christensenItNeverRains2009} treated spikes as count events arising from latent system stress, and \citet{christensenForecastingSpikesElectricity2012} produced one-step-ahead probabilities from an autoregressive conditional hazard. \citet{eichlerModelsShortTermForecasting2014} compared dynamic binary response models with that hazard model. Later studies addressed links between regions \citep{clementsModellingInterregionalLinks2015,mannerModelingForecastingMultivariate2016}, day-ahead probabilities of extreme prices \citep{liuForecastingOccurrenceExtreme2022} and clustering across days \citep{luElectricityPriceSpike2023}. These models forecast from numeric inputs. Their thresholds and horizons vary, so results do not transfer directly, and most predate five-minute settlement \citep{aemcFiveMinuteSettlement2021}.

\subsection{Evaluating forecasts of rare events}

Strictly proper scoring rules are optimised in expectation only when the forecast equals the true probability \citep{gneitingStrictlyProperScoring2007}. The Brier score \citep{brierVerificationForecastsExpressed1950} decomposes into reliability, resolution and uncertainty \citep{murphyNewVectorPartition1973}. This matters when an input is added and the model refitted. A monotone recalibration preserves the ordering of cases up to ties and can lower the reliability term \citep{dimitriadisStableReliabilityDiagrams2021}, so the refitted model may score better partly because it is recalibrated. A comparison meant to isolate the input therefore needs a control refitted in the same way without it.

Spikes also cluster, so the number of independent events is far smaller than the number of scored forecasts; resampling whole blocks of observations keeps the dependence within each unit \citep{kunschJackknifeBootstrapGeneral1989,efronIntroductionBootstrap1993}. \citet{lagoForecastingDayaheadElectricity2021} observed that new price forecasting methods were often evaluated on private datasets, on short test periods from a single market, against weak benchmarks and without testing whether differences could be due to chance. Rolling-origin evaluation mimics the way a forecast is used \citep{tashmanOutofsampleTestsForecasting2000}, but repeated analysis of one test period lets choices adapt to its results. Preregistration fixes hypotheses and analysis plans in writing before the outcome data are seen \citep{nosekPreregistrationRevolution2018}. We found no preregistered price forecasting study in the literature reviewed here.

\subsection{LLMs and market text}

One line of work uses the LLM as the forecaster \citep{gruverLargeLanguageModels2023,jinTimeLLMTimeSeries2024}. Ablations of three such forecasters found that removing the language model did not reduce accuracy \citep{tanLanguageModelsActually2024}, and the best pretrained time-series models matched, but did not consistently exceed, simple models on European day-ahead prices \citep{hornekBenchmarkingPreTrainedTime2025}.

A second line, which we follow, passes the output of an LLM reader to a numeric forecaster. Crude oil studies had added text features to numeric forecasters before LLMs \citep{liTextbasedCrudeOil2019,baiCrudeOilPrice2022}. For the NEM, \citet{huangLLMEnhancedShortTermElectricity2025} aggregated LLM-classified news into daily features lagged by one day. In a Diebold--Mariano test on 22 days of May 2024, the difference from the forecaster without news was not distinguishable from zero over all intervals, but was distinguishable in intervals with prices above \$100/MWh, a subset defined by realised prices. \citet{luLargeLanguageModelBased2025} combined a fine-tuned LLM that forecasts generator bids with a second LLM that scores news sentiment, and reported lower errors than AEMO's half-hourly forecast issued at 04:00 for the full system in May 2024; the news was examined only in ablations without a test. \citet{xueLLMEnhancedFeatureEngineering2025} and \citet{wangElectricityMarketPrice2025} added LLM scores of weather reports and of news, and neither tested the increment. \citet{biNSWEPNewsNewsAugmentedBenchmark2025}, a preprint, paired New South Wales prices with GPT-4o summaries of the news, and its text baselines used a bag-of-words representation of those summaries.

Table~\ref{tab:related-compare} compares these studies.

Two further points matter here. Dictionaries and rules can map text to categories \citep{gentzkowTextData2019}, and each notice carries a type, an effective date and a last-changed time in structured fields \citep{aemoMMSMarketNotice}, so an LLM extractor should be judged against rules and not only against a model without text. Timing also matters. A text dated before the origin may not have been visible at the origin, and an LLM may carry knowledge of the evaluation period. Back-tests can be biased when training and test periods overlap \citep{glassermanAssessingLookaheadBias2023}, and a preprint reports that LLMs recall economic values from before their knowledge cut-off even when instructed not to \citep{lopezliraMemorizationProblemCan2025}. In the present study the holdout postdates the release of the LLM, but the development folds do not.

Relative to the studies reviewed here, the question we address is therefore narrow: for a target defined by spike onset and scored as a probability, does an LLM add information beyond a strong numeric model, under an as-of rule for market text, against a rule-based baseline, and on a test period fixed before it was examined?

\begin{table}[!htbp]
  \centering\footnotesize
  \setlength{\tabcolsep}{2.5pt}
  \caption{Spike models and studies that add LLM-processed text to price forecasts, among the studies reviewed here. The columns are the design requirements of the present study, so its own row meets them by construction. ``Increment tested'' means a statistical test of the forecast with and without the text. ``Day'' marks text aligned by calendar day or a fixed daily lag. n.r., not reported; --, not applicable or not assessed.}
  \label{tab:related-compare}
  \begin{tabular}{>{\raggedright\arraybackslash}p{2.1cm}>{\raggedright\arraybackslash}p{2.4cm}c>{\raggedright\arraybackslash}p{1.5cm}>{\raggedright\arraybackslash}p{2.2cm}>{\raggedright\arraybackslash}p{1.3cm}>{\raggedright\arraybackslash}p{1.3cm}>{\raggedright\arraybackslash}p{1.3cm}>{\raggedright\arraybackslash}p{1.15cm}}
    \toprule
    Study & Target and threshold & NEM & Text source & Test period & Text timing & Non-LLM text baseline & Increment tested & Fixed before test data \\
    \midrule
    \citet{christensenForecastingSpikesElectricity2012} & Spike probability; 100~AUD/MWh & Yes & None & July to September 2007 & -- & -- & -- & -- \\
    \citet{eichlerModelsShortTermForecasting2014}$^{a}$ & Spike probability; 100 and 300~AUD/MWh & Yes & None & July to September 2007; 2011 & -- & -- & -- & -- \\
    \citet{liuForecastingOccurrenceExtreme2022} & Extreme-price probability; 95th percentile (also 90th, 99th) & Yes & None & 25\% of the 2019 data (dates n.r.) & -- & -- & -- & -- \\
    \citet{luElectricityPriceSpike2023} & Daily spike count and probability; threshold estimated for each year & Yes & None & July 2019 to September 2021, NSW & -- & -- & -- & -- \\
    \citet{hornekBenchmarkingPreTrainedTime2025} & Day-ahead price & No & None & 2024 & -- & -- & -- & -- \\
    \midrule
    \citet{huangLLMEnhancedShortTermElectricity2025} & Five-minute price & Yes & News articles & 22 days of May 2024 & Day & No & Yes & n.r. \\
    \citet{luLargeLanguageModelBased2025} & Price; spike detection & Yes & News sentiment & May 2024 (95 spikes above 600~AUD/MWh) & Day & No & No & n.r. \\
    \citet{xueLLMEnhancedFeatureEngineering2025} & Daily realised volatility & Yes & Weather reports & 2017 to 2019 & Not stated & No & No & n.r. \\
    \citet{wangElectricityMarketPrice2025} & Hourly price & No & News impact scores & 2024 (split n.r.) & Day & No & No & n.r. \\
    \citet{biNSWEPNewsNewsAugmentedBenchmark2025} (preprint) & Next-day price & Yes & LLM summaries of news & 2020 to 2024, nested & Day & On LLM summaries & No & n.r. \\
    \midrule
    Present study & Onset probability; 600~AUD/MWh & Yes & AEMO market notices & 15 August to 17 October 2026 (5 onsets) & Measured delay (proxy) & Yes & Yes & Yes (frozen locally) \\
    \bottomrule
  \end{tabular}

  \smallskip
  \parbox{\textwidth}{\footnotesize $^{a}$ Test periods and thresholds from the working-paper version, ``Modeling spike occurrences in electricity spot prices for forecasting'', METEOR Research Memorandum RM/12/029, Maastricht University (2012); the full text of the journal version was not available to us.}
\end{table}

\section{Data and as-of rules}\label{sec:data}

All data are public market data published by AEMO. We use all five NEM regions (NSW1, QLD1, SA1, TAS1 and VIC1). Table~\ref{tab:data-periods} lists the sources and periods.

\paragraph{Numeric data.} Numeric inputs for the development back-test and the final models come from the monthly archives of the Market Management System Data Model (MMSDM). These hold final, revised values and do not record whether those values match what was visible in real time. The frozen opening script downloaded the holdout numeric data; no numeric data from the evaluation window had been downloaded when the protocol was frozen.

Two AEMO price forecasts serve as inputs and benchmarks, not as objects of correction. P5MIN runs every five minutes and covers the next hour \citep{aemoMMSP5MIN}. PD runs every half hour and covers each period up to the end of the last trading day for which bids have closed \citep{aemoPredispatchProcedure2026}. The operator produces both with access to all bids, constraints and demand forecasts.

\paragraph{Market notices.} Each market notice is a row of the MMSDM table MARKETNOTICEDATA, with an effective date, a type, a last-changed timestamp and free text of up to 2,000 characters \citep{aemoMMSMarketNotice}. We removed four routine types from the training notices (prices subject to review, prices unchanged, settlements residue and market systems).

From August 2024 the MMSDM no longer contains this table, and the NEMweb directory of current notices keeps only about the most recent 60 days. The data used in this study therefore contain no notice text for August 2024 to mid-August 2026. This gap is the main constraint on the design: the text models learn from notices up to July 2024 and are applied to notices from August 2026.

Holdout notices are restricted to the 15 types retained in the training data. We fixed this restriction after seeing the counts of notice types among the captured holdout files; a sensitivity analysis without it is part of the frozen protocol. In the holdout range 24 notice identifiers (about 2.8\%) are absent from our capture. A missing update or cancellation can leave an earlier state, such as an active lack-of-reserve level, in place until its stated end.

\begin{table}[!htbp]
  \centering\small
  \caption{Data sources and periods of the study. All periods before the holdout are development data.}
  \label{tab:data-periods}
  \begin{tabular}{>{\raggedright\arraybackslash}p{3.5cm}>{\raggedright\arraybackslash}p{7.2cm}>{\raggedright\arraybackslash}p{4.0cm}}
    \toprule
    Item & Source and coverage & Use \\
    \midrule
    Numeric data, development & MMSDM dispatch, regional, P5MIN and PD tables, October 2021 to July 2026 & Inputs and labels of the back-test and the final models \\
    Numeric data, holdout & NEMweb archive and current directories, downloaded by the frozen opening script & Holdout inputs and labels \\
    Training notices & MMSDM MARKETNOTICEDATA, October 2021 to July 2024; 21,954 notices, 14,019 after removing four routine types & Notice features and text increments \\
    Holdout notices & NEMweb current directory from 14 August 2026; backfill, then polling every 120 seconds, from shortly after midnight AEST on 13 October 2026; 15 retained types & Holdout notice features \\
    No notice text used & August 2024 to mid-August 2026 & None \\
    \midrule
    Development back-test & Five regions; rolling origins July 2022 to July 2024 in nine folds; 365,622 scored origins at 60 minutes, 365,313 at 24 hours & Development out-of-sample evidence \\
    Final models & Numeric part trained to July 2026, text part to July 2024 & Holdout forecasts \\
    Holdout & Five regions; origins 15 August to 17 October 2026 on a 15-minute grid; 64 days and 30,720 origin--region pairs at each horizon & Confirmatory test, opened once \\
    \bottomrule
  \end{tabular}
\end{table}

\paragraph{Clocks and visibility.} All times are in Australian Eastern Standard Time (AEST, UTC+10), without daylight saving. An AEMO interval timestamp marks the end of the interval \citep{aemoMMSDispatchPrice}. An input may be used at origin $t$ only if it was available at $t$. For P5MIN and PD we use the latest run whose last-changed time is no later than $t$. A training notice is treated as visible at
\begin{equation}
  v = \max(\text{effective date}, \text{last changed}) + \Delta ,
  \label{eq:visible}
\end{equation}
a holdout notice at its creation time plus $\Delta$, and a notice may be used at $t$ only if $v \le t$.

A rule in the first frozen version sets $\Delta = \max(15, \lceil p_{99} \rceil_5)$ minutes, where $p_{99}$ is the 99th percentile of the measured publication delay and $\lceil \cdot \rceil_5$ rounds up to five minutes. The delay runs from a notice's stated creation time to its first appearance in the current directory, for notices created after live polling began. The 43 notices observed live between 13 and 17 October 2026 gave a median of 4.99 minutes, a 99th percentile of 24.52 and a maximum of 34.45, so $\Delta = 25$. Among the 27 measured notices of retained types the 99th percentile was 28.3 minutes, which would round to 30; the frozen rule uses all measured notices. Seven notices captured during polling interruptions were excluded under a rule in the frozen protocol, three of them with an apparent delay of 93.3 minutes. One delay exceeded $\Delta$, and the exact one-sided 95\% upper bound on the probability of exceeding it is 0.1056 \citep{clopperUseConfidenceFiducial1934}.

This is a proxy rule, not a record of when each notice could be read. The archive gives no time of first publication, so the delay measured in October 2026 is assumed to hold for 2021 to 2024. Delays for 15 August to 12 October 2026 were not measured. All five holdout onsets fall in this unmeasured period, so no delay was measured on an event day. A holdout notice published 25 to 35 minutes after creation would be treated as visible up to about 9.5 minutes early. Models with $\Delta = 60$ bound this effect for H1 and H2, but no such sensitivity analysis exists for H3.

\paragraph{Separation of development and holdout data.} A ledger of the first author's earlier data use found no work that had read market values from the holdout period. Two earlier checks touched its metadata: header-only requests for daily dispatch files dated up to 21 August 2026, and a directory listing of file names. Neither reveals prices or labels. During a rehearsal, 48 PD runs published on 1 August 2026, in the look-back period 14 days before the first origin, were parsed mechanically; they entered no feature and were not inspected. The first author, who directed all analyses, and the analysis pipeline read no holdout market value before the final freeze. Holdout notices were processed only mechanically: their publication times and type counts were used as described above, and only the counts, parse rates and hashes of the outputs were checked. We were not blind to the notices of October 2021 to July 2024, which the first author's earlier work had used for NSW1.

\section{Targets}\label{sec:targets}

Let $p_\tau$ be the unrounded regional price of the five-minute dispatch interval ending at time $\tau$, with times in minutes. An interval $\tau$ is a spike onset if $p_\tau > 600$~AUD/MWh and the three preceding prices are at or below 600. The episode that begins at onset $S$ ends at the first time $R > S$ at which three consecutive prices are at or below the threshold,
\begin{equation}
  R = \min\{\tau > S : p_\tau \le 600,\; p_{\tau+5} \le 600,\; p_{\tau+10} \le 600\}.
  \label{eq:recovery}
\end{equation}
For a forecast origin $t$, let $S^\ast(t)$ be the first onset after $t$. The two binary targets are
\begin{equation}
  y_t = \mathbf{1}\{\, t < S^\ast(t) \le t + 60 \,\},
  \qquad
  y^{24}_t = \mathbf{1}\{\, t < S^\ast(t) \le t + 24\,\mathrm{h} \,\}.
  \label{eq:onset-label}
\end{equation}
An onset at $t$ itself is not a target, and a later onset counts only after the earlier episode has recovered. A positive origin is one with $y = 1$ at the horizon concerned. At 60 minutes an origin is scored only if $t + 60$ is strictly earlier than the end of the data it belongs to; at 24 hours only if those data cover its whole window up to $t + 24$~h.

We score all origins on a 15-minute grid, including those inside an episode. We exclude origins within 60 minutes either side of an interval in which the administered price cap or floor applied or the market was suspended, because those prices are set administratively. At 24 hours the exclusion extends to $[t-60, t+24\,\mathrm{h}]$. The first frozen version excluded any non-zero administered-pricing flag; we keep that literal rule as a sensitivity analysis.

The threshold is the administered price cap that applies after a sustained period of high prices. It was raised from 300 to 600~AUD/MWh from 1 December 2022 \citep{aemcAmendingAPC2022} and retained for 1 July 2025 to 30 June 2028 \citep{aemcMarketPriceSettings2023}, so the first development fold, and most of the second, precede the raise. We inherited the threshold from the first author's earlier spike-warning work and do not test whether another threshold would change the conclusions. The frozen protocol also names timing, duration (the minutes above 600~AUD/MWh within the next hour, not the episode length) and magnitude as secondary outcomes. With five holdout onsets they cannot be interpreted; Section~\ref{sec:holdout} gives the point values computed.

\section{Feature sets and forecaster}\label{sec:features}

\begin{figure}[!htbp]
  \centering
  \begin{tikzpicture}[font=\footnotesize, node distance=0.45cm and 0.5cm,
  box/.style={draw=black!60, rounded corners=2pt, align=center, text width=3.3cm, minimum height=1.0cm, inner sep=3pt, fill=white},
  llm/.style={box, fill=black!10},
  arr/.style={-{Stealth[length=2mm]}, black!65}]
\node[box] (num) {Numeric market data\\ prices, demand, P5MIN, PD\\ (as of the origin)};
\node[box, below=of num] (not) {Market notices\\ visible at $v$ (Eq.~\eqref{eq:visible})};
\node[llm, right=of not] (ext) {LLM, once per notice:\\ structured fields (L1)\\ and embedding (L2)};
\node[box, below=of not] (rule) {Rule-based notice\\ features (B)};
\node[box, right=of ext] (agg) {Aggregate to region\\ $\times$ origin\\ (decayed counts, LOR state)};
\node[box, above=of agg] (A) {Numeric model A\\ (LightGBM, cross-fitted)};
\node[box, right=of agg] (inc) {Text increment on\\ A's logit (offset);\\ A0 = no text};
\node[box, above=of inc] (prob) {Onset probability,\\ 60 min and 24 h};
\node[box, below=of inc] (eval) {Development back-test\\ (9 folds) $\to$ final freeze\\ $\to$ holdout, opened once};
\draw[arr] (num) -- (A);
\draw[arr] (not) -- (ext);
\draw[arr] (not) -- (rule);
\draw[arr] (ext) -- (agg);
\draw[arr] (rule.east) -| (agg.south);
\draw[arr] (A) -- (inc);
\draw[arr] (agg) -- (inc);
\draw[arr] (inc) -- (prob);
\draw[arr] (inc) -- (eval);
\end{tikzpicture}
  \caption{Pipeline of the study. The LLM (shaded) reads each notice once, after the notice is deemed visible; all probabilities come from LightGBM. The text features enter as an increment on the logit of the numeric model A, and A0 is the same increment without text. Development results come from a nine-fold rolling back-test; the confirmatory test uses a holdout opened once after the protocol was frozen.}
  \label{fig:pipeline}
\end{figure}
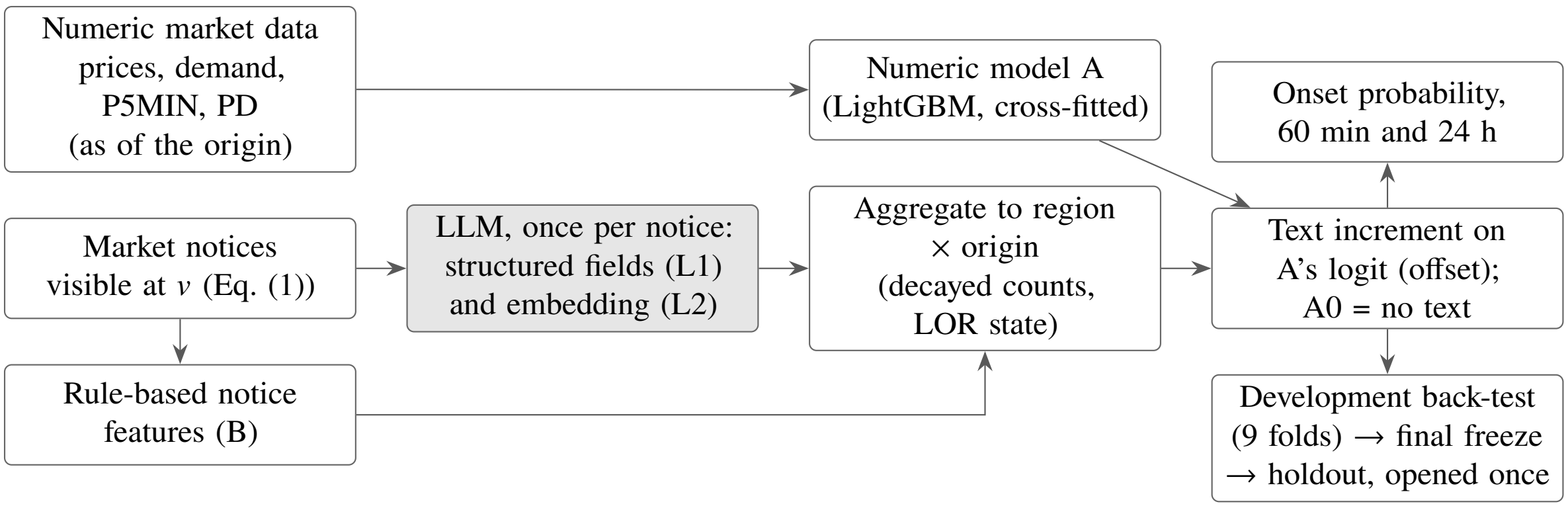

The LLM only reads market notices, once per notice, and a local gradient-boosting model produces every probability (Figure~\ref{fig:pipeline}). The extracted features can therefore be frozen and hashed before any model is fitted, the forecaster can be trained on hundreds of thousands of origins, and a difference in score between feature sets cannot come from a different learner.

\paragraph{Feature sets.} \emph{A} contains 34 numeric features: current and recent prices, demand, available generation and reserve margin, net interchange, demand forecasts, P5MIN and PD summaries, prices in other regions, calendar terms and the region. The same 34 features are used at both horizons. The P5MIN and PD summaries look no further than the next hour, and no reserve data from AEMO's projected assessment of system adequacy (PASA) \citep{aemoSTPASA2025} enter A. At 24 hours A therefore lacks AEMO's forward price and reserve forecasts, and some forecast lack-of-reserve notices themselves report PASA results.

\emph{B} adds notice features produced by fixed rules. Regular expressions map each notice to regions and extract the first lack-of-reserve (LOR) level stated, the largest megawatt figure and cancellations. Notices are counted by type with an exponential decay of half-life three hours.

\emph{L1} adds the same kinds of features produced by the LLM. We ran a community text-only build of Qwen3.5-9B \citep{qwenTeamQwen35ModelCard2026} at a fixed revision, with a fixed prompt, greedy decoding and at most 160 new tokens. It returned eight fields per notice: an event type from 12 classes, the regions, the LOR level, the megawatt figure, the effective start and end times, the status and any referenced notice. These are aggregated to each region and origin as decayed counts by event type and status, the currently active LOR level and megawatts, and the number of recent notices whose stated event window overlaps $(t, t+60]$. We did not validate the extracted fields against human labels, and there is no public evidence on their accuracy. In an exploratory check on the 14,019 training notices, made before the final freeze, the LLM agreed with the fixed rules on the LOR level of 99.5\% of reserve notices, on the region for 99.97\% of notices where the rules found one region, and on cancellation for 99.7\%. Where both gave a megawatt figure (3,923 notices), the LLM's value matched a figure stated in the notice in only 35\%. Agreement with rules is not accuracy.

\emph{L2} adds an embedding of each notice: the mean of the last hidden layer over at most 512 tokens, reduced by principal component analysis to 16 components and averaged over recent notices with the same decay. An amendment fixed the number of components after a development comparison of 8, 16 and 32. The components were fitted once on all training notices, so the features of each development fold use directions estimated partly from later notices. This affects only L2.

\paragraph{Forecaster.} We chose the forecaster on set A before any text feature was used. Before two later corrections to the sample and to one feature, LightGBM \citep{keLightGBMHighlyEfficient2017} had a lower sample-weighted Brier score on the last four development folds than a discrete-time logistic hazard \citep{singerItsTimeUsing1993} (0.004847 against 0.005327). It uses 31 leaves, at least 200 rows per leaf and a learning rate of 0.03, with early stopping on the last 10\% of each training set.

The training set, ordered by time, is split into five consecutive blocks, and A is the bag of five models, each fitted on four blocks. Each training row receives the logit of the model that did not see its block, which may have been fitted on later training data. Evaluation forecasts use the average logit of the five models. With 24-hour labels, the block boundaries and the early-stopping split have 23 hours of label overlap, which uses no data from the evaluation period.

\paragraph{Text increment and control.} Text enters through an increment model fitted with A's out-of-fold logit as a fixed offset. Its features are the text features of the set, the reserve margin, the hour and the region. Because this model can also recalibrate A, we fit a control increment of the same form with no text features, denoted \emph{A0}. Comparisons against A0 are intended to separate text from recalibration, although the two increments may still recalibrate A differently, as the final 24-hour increments do (Section~\ref{sec:development}). Each specification is averaged over five seeds.

\paragraph{Final models and references.} The final numeric model was trained on data up to July 2026, and the text increments on notices up to July 2024. In that final refit, on the whole text window of October 2021 to July 2024, early stopping kept a single tree in the 60-minute increment for every text set and seed.

The references are a climatological rate by region and hour, and the P5MIN and PD price forecasts converted to probabilities by isotonic regression \citep{zadroznyTransformingClassifierScores2002}. The 24-hour P5MIN and PD references are AEMO's next-hour forecasts mapped in this way, not AEMO 24-hour forecasts.

\section{Evaluation design}\label{sec:evaluation}

\paragraph{Scores.} Probability forecasts $q_t$ are scored with the strictly proper Brier score \citep{brierVerificationForecastsExpressed1950,gneitingStrictlyProperScoring2007}, and comparisons use the relative skill
\begin{equation}
  \mathrm{BS} = \frac{1}{n}\sum_t (q_t - y_t)^2,
  \qquad
  \mathrm{BSS}_{\mathrm{rel}} = 1 - \frac{\mathrm{BS}_{\text{candidate}}}{\mathrm{BS}_{\text{reference}}}.
  \label{eq:bss}
\end{equation}
Over fixed probability bins the score decomposes as $\mathrm{BS} = \mathrm{REL} - \mathrm{RES} + \mathrm{UNC}$ \citep{murphyNewVectorPartition1973}, plus a within-bin term \citep{stephensonTwoExtraComponents2008}. The fixed bins have edges 0, 0.001, 0.002, 0.005, 0.01, 0.02, 0.05, 0.1, 0.2, 0.5 and 1.

\paragraph{Development back-test.} Development evidence comes from a rolling-origin back-test \citep{tashmanOutofsampleTestsForecasting2000} over nine folds from July 2022 to July 2024: calendar quarters to June 2024, plus July 2024. Models are trained on origins from October 2021 up to one hour before each fold, so every 60-minute training label closes before the fold begins. At 24 hours the same cut means that the labels of the last training origins overlap the first 23 hours of the fold. We selected hyperparameters on the last four folds, which start in October 2023, and report the other five folds separately at 60 minutes; at 24 hours we report pooled values and the range across folds. We call these results development out-of-sample. They are not confirmatory.

\begin{table}[!htbp]
  \centering\small
  \caption{Confirmatory hypotheses. All are one-sided tests of a lower Brier score on the holdout, five regions pooled. $\alpha$ is the initial level under the graphical Holm procedure; H2 starts at zero and receives the level of H1 only if H1 is established, and levels pass between hypotheses as described in the text.}
  \label{tab:hypotheses}
  \begin{tabular}{>{\raggedright\arraybackslash}p{0.8cm}>{\raggedright\arraybackslash}p{4.8cm}>{\raggedright\arraybackslash}p{1.5cm}>{\raggedright\arraybackslash}p{1.0cm}>{\raggedright\arraybackslash}p{5.6cm}}
    \toprule
    & Comparison & Horizon & $\alpha$ & Statistic \\
    \midrule
    H1 & L1 against A & 60 min & 0.025 & Day-cluster bootstrap \\
    H2 & L1 against B, tested only after H1 is established & 60 min & 0 & Day-cluster bootstrap \\
    H3 & L1 against A0 & 24 h & 0.025 & Bootstrap over non-overlapping three-day blocks \\
    \bottomrule
  \end{tabular}
\end{table}

\paragraph{Hypotheses.} Table~\ref{tab:hypotheses} lists the hypotheses. Because the final 60-minute increments are single trees, H1 and H2 compare nearly identical forecasts, and the outcome recorded as expected before opening was a label inside the equivalence band. We kept H1 and H2 as fixed in the first version and recorded this expectation, rather than changing them after the single-tree result was known.

We added H3 on 16 October 2026, after the development back-test had shown a favourable 24-hour difference for L1, so that value is a basis for the hypothesis, not evidence for it. H3 uses A0 as reference so that recalibration available to the text-free increment is not counted as text information. The separation is imperfect: the final 24-hour A0 stopped at one tree and the L1 increment at 8 to 16, so the two differ in complexity. Establishing H3 would show that the frozen L1 pipeline had a lower 24-hour Brier score than the frozen text-free increment. Because the text increment shifted forecasts even when no notice was visible, such a difference may still include recalibration. And because H3 does not involve set B, establishing it would not show that LLM extraction adds information beyond fixed rules.

\paragraph{Multiplicity and inference.} We use a graphical Holm procedure \citep{bretzGraphicalApproachSequentially2009}. A hypothesis is established, meaning that its null hypothesis of no reduction in the Brier score is rejected, only if its one-sided p-value is below its current level and its relative skill is not negative. When H1 is established its level passes to H2; when H2 is established its level passes to H3; when H3 is established its level passes to H1, or to H2 if H1 has already been established and H2 has not. This holds the family-wise error rate at 0.05. The bootstrap resamples AEST days, keeping the regions of a day together, with 10,000 resamples on the holdout and 2,000 in development. H3 uses three-day blocks because 24-hour labels overlap across neighbouring days.

\paragraph{Smallest effect and labels.} The smallest effect of interest is a relative Brier reduction of 2\%, and equivalence is assessed with two one-sided tests against bounds of $\pm$2\% \citep{schuirmannComparisonTwoOneSided1987,lakensEquivalenceTests2017}. The 2\% bound is a judgement fixed in the first version; it was not derived from decision value.

Each comparison also receives one of six descriptive labels, assigned in this order. \emph{Superior but below the smallest effect of interest}: one-sided p-value below 0.05, skill not negative and 90\% interval within $\pm$2\%. \emph{Superior}: p-value below 0.05, skill not negative and 90\% interval not within $\pm$2\%. \emph{Equivalent but slightly worse}: 90\% interval within $\pm$2\% and 95\% upper bound below zero. \emph{Equivalent}: 90\% interval within $\pm$2\%. \emph{Inferior}: 95\% upper bound below zero. \emph{Inconclusive}: anything else. The labels describe statistical evidence at the 0.05 level and do not replace the confirmatory decisions.

\paragraph{Event-level alarms.} Alarm thresholds were frozen on development out-of-sample forecasts to allow at most two false alarms per region per week. An alarm is raised at an origin whose probability reaches the threshold when none has done so in the previous 60 minutes. An onset counts as detected if an alarm falls on one of the positive origins before it; any other alarm is a false alarm.

\paragraph{Prespecified descriptive analyses for H3.} An amendment written before the final freeze, after development results and a pre-holdout rehearsal had been seen, and before any holdout market data had been read, attached two analyses to H3. They enter no error-rate control. First, origins are split by whether the 12 notice-count features of L1 sum to more than zero, and L1 is compared with A0 within each stratum. Second, the 24-hour scores of L1, A0 and A are decomposed on the fixed bins.

Two rules fix the wording, both conditional on L1 having the lower Brier score on all origins. If the stratum without notices carries more than half of the size-weighted reduction, the report must state that the improvement occurs mainly at origins without notices and cannot be attributed to notice content. If the reduction in the reliability term from A0 to L1 exceeds the increase in the resolution term, the report must state that the improvement comes mainly from recalibration and cannot be attributed to notice content.

\paragraph{The two freezes.} Preregistration separates prediction from postdiction \citep{nosekPreregistrationRevolution2018}, but our evidence for it is local. We froze the protocol on our own machines and hashed it. We did not deposit it in a public registry, and no third party timestamped it. Publishing the hashes shows that files are unchanged, not when they were frozen.

The first version was frozen at 01:00:19 AEST on 13 October 2026, before any forecasting model had been fitted. The second was frozen at 01:53:11 AEST on 18 October 2026, after 44 dated amendments, many of them prompted by development results or by the review of the opening procedure. It fixes $\Delta$, the final models, the hypotheses and the scripts. 

The holdout is retrospective. We replaced a planned prospective window with the most recent period not used in the first author's earlier work, so that it could be evaluated before a fixed deadline. All five onsets had occurred, and their prices were public, before the first freeze. The protection is that the first author and the analysis pipeline read no holdout market value before the final freeze, and this rests on our own records. The frozen script opened the holdout once on 20 October 2026, producing all forecasts before reading any label. Every analysis added after opening is labelled post hoc. The frozen sensitivity analyses are listed with their results in Table~\ref{tab:hold-sens}, and Appendix~\ref{app:deviations} lists the deviations.

\paragraph{Expected power.} Before the final freeze we replayed development forecasts in 64-day windows. For the 26 windows with 80 to 160 positive origins, the number then expected, the median minimum detectable effect with 80\% power at a one-sided level of 0.05 was 2.07\% for H1 and 4.89\% for H3, against a development estimate for H3 of $+$1.65\% [0.54, 2.74]. At level 0.025, H3 met the criterion in 27\% of the windows and was inconclusive in 65\%. At most five windows are mutually non-overlapping, and H3 was chosen after its development signal had been seen, so these figures are imprecise and an upper bound on its power. After the post-freeze 24-hour refit (Section~\ref{sec:development}), H3 met the criterion in 23\% of the windows and was inconclusive in 77\%, and the median detectable effect lay between 2.8\% and 4.9\% in the two computations. Section~\ref{sec:holdout} reports how far the holdout fell short of these windows.

\section{Development back-test}\label{sec:development}

In the development back-test, market notice features did not lower the 60-minute Brier score of a numeric model that already includes AEMO's forecasts. At 24 hours the text sets lowered the Brier score, but rule-based features shared the reduction, and its attribution to notice content was not established. All results here are development out-of-sample. Every forecast came from models trained on earlier origins, but the hyperparameters were selected on the last four folds, and many of the 44 protocol amendments were written after these results had been seen. Values use the final 25-minute visibility delay and five seeds. Comparisons are relative Brier skill in percent with 95\% intervals; positive values favour the first set.

The back-test covered 365,622 origins in five regions over nine folds from July 2022 to July 2024, of which 3,631 were followed by an onset within 60 minutes. At 24 hours it gave 365,313 scored origins with 51,198 positives, although overlapping 24-hour windows make neighbouring labels highly dependent.

\subsection{Sixty-minute onset}

\begin{figure}[!htbp]
  \centering
  \includegraphics[width=\textwidth]{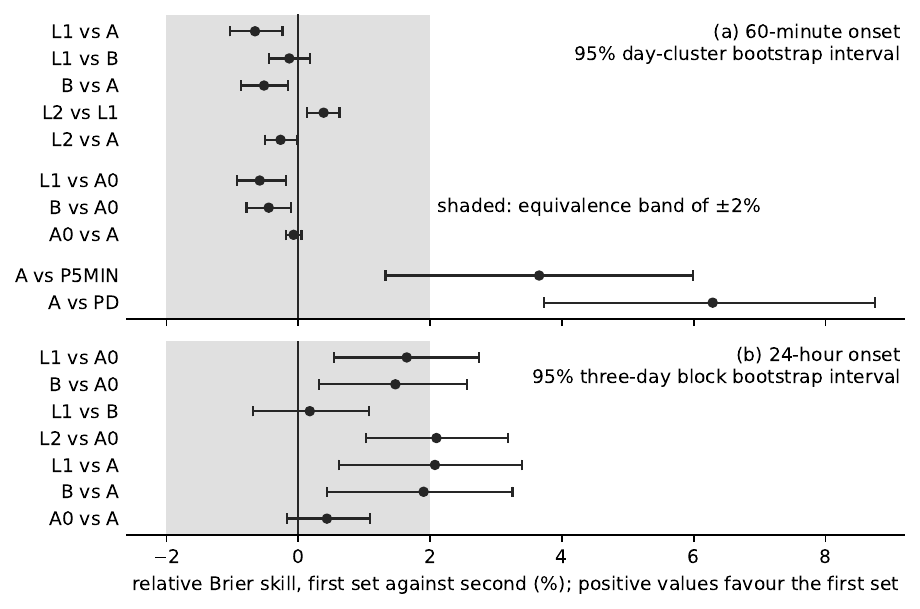}
  \caption{Development back-test: relative Brier skill (\%) of the first feature set against the second, for the 60-minute onset (panel a, five regions, all nine folds pooled) and the 24-hour onset (panel b, five regions). These are development out-of-sample results, not holdout results. Points are estimates and bars are 95\% intervals, from a day-cluster bootstrap in panel a and a three-day block bootstrap in panel b. The vertical line marks zero and the shaded area the equivalence band of $\pm$2\%. A0 is the text-free control increment; P5MIN and PD are AEMO's forecasts converted to probabilities. The 24-hour values are from the original back-test, before the post-freeze refit.}
  \label{fig:dev-forest}
\end{figure}

At 60 minutes every comparison among the text sets fell inside the equivalence band of $\pm$2\% (Table~\ref{tab:dev-60} and Figure~\ref{fig:dev-forest}). L1 against A was $-$0.66\% [$-$1.04, $-$0.24], labelled equivalent but slightly worse, and B against A was $-$0.52\% [$-$0.87, $-$0.15]. L1 and B were equivalent ($-$0.14\% [$-$0.45, 0.18]). The text-free control A0 did not differ from A ($-$0.07\% [$-$0.19, 0.05]), so the small loss of L1 belongs to the text features rather than to the increment model. The same pattern held on the five folds not used for hyperparameter selection. These folds do not describe the near-degenerate model tested by H1: the median number of trees kept by the L1 increment across seeds ranged from 1 to 71 across folds, against one tree in the final model.

\begin{table}[!htbp]
  \centering\small
  \caption{Development back-test, 60-minute onset: relative Brier skill (\%) of the first feature set against the second, with day-cluster bootstrap 95\% intervals, for all nine folds (pooled) and for the five folds not used for hyperparameter selection (Non-selection). Positive values favour the first set. A0 is the text-free control increment; P5MIN and PD are AEMO's forecasts converted to probabilities by isotonic regression. Labels are those of Section~\ref{sec:evaluation}.}
  \label{tab:dev-60}
  \begin{tabular}{lrl>{\raggedright\arraybackslash}p{3.2cm}rl}
\toprule
Comparison & $\mathrm{BSS}_{\mathrm{rel}}$ & 95\% CI & Label (pooled) & Non-selection & 95\% CI \\
\midrule
L1 vs A & $-$0.66 & [$-$1.04, $-$0.24] & equivalent but slightly worse & $-$0.63 & [$-$1.10, $-$0.11] \\
L1 vs B & $-$0.14 & [$-$0.45, 0.18] & equivalent & $-$0.14 & [$-$0.52, 0.25] \\
B vs A & $-$0.52 & [$-$0.87, $-$0.15] & equivalent but slightly worse & $-$0.49 & [$-$0.94, $-$0.05] \\
L2 vs L1 & +0.38 & [0.13, 0.63] & superior but below the smallest effect of interest & +0.33 & [0.02, 0.63] \\
L2 vs A & $-$0.27 & [$-$0.50, $-$0.02] & equivalent but slightly worse & $-$0.30 & [$-$0.59, 0.03] \\
\midrule
L1 vs A0 & $-$0.58 & [$-$0.93, $-$0.18] & equivalent but slightly worse & $-$0.59 & [$-$1.02, $-$0.09] \\
B vs A0 & $-$0.45 & [$-$0.78, $-$0.11] & equivalent but slightly worse & $-$0.44 & [$-$0.88, $-$0.03] \\
A0 vs A & $-$0.07 & [$-$0.19, 0.05] & equivalent & $-$0.05 & [$-$0.19, 0.09] \\
\midrule
A vs P5MIN & +3.66 & [1.32, 5.99] & superior & +1.98 & [$-$0.79, 4.76] \\
A vs PD & +6.29 & [3.72, 8.76] & superior & +5.45 & [2.55, 8.14] \\
\bottomrule
\end{tabular}

\end{table}

A and L1 had the same area under the ROC curve (0.941), but L1 had a larger reliability term and a smaller resolution term (Table~\ref{tab:dev-murphy}).

\begin{table}[!htbp]
  \centering\small
  \caption{Murphy decomposition of the 60-minute Brier score in the development back-test (five regions, nine folds, five seeds), with the area under the ROC curve (AUC) and under the precision--recall curve (PR-AUC). REL is reliability (lower is favourable) and RES resolution (higher is favourable).}
  \label{tab:dev-murphy}
  \begin{tabular}{lrrrrr}
\toprule
Model & Brier $\times 10^{3}$ & REL $\times 10^{5}$ & RES $\times 10^{5}$ & AUC & PR-AUC \\
\midrule
A & 8.270 & 3.77 & 156.1 & 0.941 & 0.267 \\
B & 8.313 & 4.62 & 152.2 & 0.942 & 0.262 \\
L1 & 8.324 & 5.01 & 151.7 & 0.941 & 0.260 \\
L2 & 8.292 & 3.89 & 155.2 & 0.942 & 0.264 \\
A0 & 8.276 & 3.68 & 155.1 & 0.940 & 0.266 \\
P5MIN & 8.584 & 5.77 & 128.3 & 0.879 & 0.192 \\
PD & 8.825 & 3.23 & 104.5 & 0.849 & 0.157 \\
\bottomrule
\end{tabular}

\end{table}

A, which uses the latest P5MIN and PD forecasts as inputs, had a higher relative skill than PD converted to a probability ($+$6.29\%) and than converted P5MIN ($+$3.66\%), although the P5MIN comparison was inconclusive on the non-selection folds ($+$1.98\% [$-$0.79, 4.76]). At 60 minutes, with an earlier 30-minute delay, L1 on real notices scored no better than L1 on notices shifted by seven days ($-$0.18\% [$-$0.54, 0.18]); this placebo was not repeated with the final delay.

\subsection{Twenty-four-hour onset}

At 24 hours, L1 against A0 was $+$1.65\% [0.54, 2.74] under the three-day block bootstrap, labelled superior because its 90\% interval extends beyond 2\%, although the point estimate is below that smallest effect of interest (Table~\ref{tab:dev-24}). H3 was added to the protocol after an earlier computation of this comparison, with a 30-minute delay, had shown a favourable signal. The value above is a basis for the hypothesis, not evidence for it. Training sets also overlapped the first hours of each fold through the 24-hour labels. A post-freeze sensitivity refit, which ended every training set 24 hours before its fold, gave $+$1.69\% [0.69, 2.69].

\begin{table}[!htbp]
  \centering\small
  \caption{Development back-test, 24-hour onset: relative Brier skill (\%) with three-day block bootstrap 95\% intervals, for the five regions and without NSW1. Positive values favour the first set; A0 is the text-free control increment. The label refers to the five-region comparison. Values are from the original back-test, the basis on which H3 was proposed, not from the post-freeze refit.}
  \label{tab:dev-24}
  \begin{tabular}{lrl>{\raggedright\arraybackslash}p{2.6cm}rl}
\toprule
Comparison & Five regions & 95\% CI & Label & Without NSW1 & 95\% CI \\
\midrule
L1 vs A0 & +1.65 & [0.54, 2.74] & superior & +0.78 & [$-$0.42, 1.97] \\
B vs A0 & +1.47 & [0.32, 2.56] & superior & +0.59 & [$-$0.79, 1.94] \\
L1 vs B & +0.18 & [$-$0.69, 1.08] & equivalent & +0.19 & [$-$0.79, 1.20] \\
L2 vs A0 & +2.10 & [1.02, 3.19] & superior & +1.53 & [0.34, 2.70] \\
L1 vs A & +2.07 & [0.62, 3.39] & superior & +1.27 & [$-$0.33, 2.62] \\
B vs A & +1.90 & [0.44, 3.25] & superior & +1.08 & [$-$0.52, 2.53] \\
A0 vs A & +0.43 & [$-$0.17, 1.09] & equivalent & +0.50 & [$-$0.08, 1.12] \\
\bottomrule
\end{tabular}

\end{table}

Several points limit this value. B reached almost the same skill against A0 ($+$1.47\% [0.32, 2.56]), and L1 and B were equivalent ($+$0.18\% [$-$0.69, 1.08]), so the signal is not specific to LLM extraction. The signal was larger in NSW1 ($+$4.87\% [1.79, 7.73]). NSW1 is the only region whose prices had been modelled with the notices of 2021 to 2024 in the first author's earlier work, so we were not blind to those notices and the NSW1 result is not independent of that work. Without NSW1, L1 against A0 was $+$0.78\% [$-$0.42, 1.97] and, in the refit, $+$1.04\% [0.04, 2.09]. By region it was $+$2.25\% [$-$0.82, 5.27] in QLD1, $-$0.09\% [$-$2.47, 2.08] in SA1, $+$1.22\% [$-$0.14, 2.80] in TAS1 and $-$0.75\% [$-$2.99, 1.74] in VIC1. Fold values ranged from $-$2.87\% to $+$5.13\%, and the first fold, before the price cap rise, holds 14,429 of the 51,198 positives. L2 against A0 was $+$2.10\% [1.02, 3.19], but L2 is outside the hypotheses, and its principal components were fitted on all training notices, including those after each fold.

The final 24-hour L1 increment kept 8 to 16 trees across seeds (median 13), whereas A0 kept one, so part of the 24-hour skill may be recalibration by the text increment. In a rehearsal window before the holdout (17 to 30 July 2026) with no visible notice, the L1 increment still moved the logit relative to A0 by a median of 0.085, and the B and L2 increments by more. The Murphy decomposition of the back-test nevertheless placed the 24-hour reduction in resolution rather than reliability (resolution 1766.6 against 1582.3, reliability 82.0 against 35.9, in units of $10^{-5}$; Table~\ref{tab:hold-murphy}). We did not run a placebo at 24 hours, so the development signal has not been tested against shifted notices.
\section{Single-use holdout evaluation}\label{sec:holdout}

The retrospective holdout was evaluated once against the hypotheses of Table~\ref{tab:hypotheses}, on origins from 15 August to 17 October 2026 in all five regions; H1 and H3 were not established and H2 was not tested. The protocol, models, visibility delay, scripts and decision rules were frozen and hashed locally on 18 October 2026 (Section~\ref{sec:evaluation}). The frozen script opened the holdout once, on 20 October 2026. It downloaded the holdout data, produced all forecasts without reading any label, recorded hashes of the forecasts, labels and scoring inputs, and only then scored them. It ran without interruption or resumption.

The expected outcomes were recorded before opening: H1 and H2 inside the equivalence band, and an inconclusive H3 as the most likely single outcome.

\subsection{Data and events}

The holdout contained 30,720 origin--region pairs at each horizon over 64 days (Table~\ref{tab:base-rates}). It had five spike onsets on three days (9, 10 and 12 October 2026) in three regions (Table~\ref{tab:hold-events}), and the 24-hour positives fell in 2 of the 22 three-day blocks, so the holdout intervals are determined largely by two blocks. The 24-hour base rate was 1.56\% against 14.0\% in development, and every model, including AEMO's forecasts, had a higher Brier score than a constant forecast at the realised base rate, which is not available in advance (post hoc description). There were 20 positive origins at 60 minutes and 480 at 24 hours. That is far fewer than in the replay windows behind the power estimate, which had 80 to 160 positives at 60 minutes (median 136), and the five onsets are fewer than the 20--34 of a design-stage projection counted differently from the frozen labels. The estimate in the frozen protocol therefore overstated the power of the holdout.

\begin{table}[!htbp]
  \centering\footnotesize
  \caption{Base rates in the development back-test and on the holdout: origins, positive origins and their share, five regions. The last column is the mean forecast probability of the numeric model A divided by the holdout base rate (post hoc). This description of the data was computed after opening.}
  \label{tab:base-rates}
  \begin{tabular}{lrrrrrr>{\raggedleft\arraybackslash}p{2.2cm}}
\toprule
 & \multicolumn{3}{c}{Development back-test} & \multicolumn{4}{c}{Holdout} \\
\cmidrule(lr){2-4}\cmidrule(lr){5-8}
Horizon & Origins & Positives & Base rate (\%) & Origins & Positives & Base rate (\%) & Mean forecast of A / base rate \\
\midrule
60 minutes & 365,622 & 3,631 & 0.993 & 30,720 & 20 & 0.0651 & 1.69 \\
24 hours & 365,313 & 51,198 & 14.0 & 30,720 & 480 & 1.56 & 2.61 \\
\bottomrule
\end{tabular}

\end{table}

\begin{table}[!htbp]
  \centering\footnotesize
  \caption{Post hoc list of the five holdout onsets (description only). Onset times are the end of the dispatch interval in AEST; the highest price, the minutes above 600~AUD/MWh within the hour and the highest 60-minute probabilities are taken over the positive origins of each onset.}
  \label{tab:hold-events}
  \resizebox{\textwidth}{!}{\begin{tabular}{ll>{\raggedleft\arraybackslash}p{1.9cm}>{\raggedleft\arraybackslash}p{1.9cm}>{\raggedleft\arraybackslash}p{1.9cm}rrrr}
\toprule
 &  &  &  &  & \multicolumn{4}{c}{Highest 60-minute probability} \\
\cmidrule(lr){6-9}
Region & Onset (AEST) & Price at onset (AUD/MWh) & Highest price within the hour (AUD/MWh) & Minutes above 600 within the hour & A & L1 & P5MIN & PD \\
\midrule
NSW1 & 9 Oct 05:00 & 992.32 & 992.32 & 35 & 0.008 & 0.008 & 0.22 & 0.03 \\
NSW1 & 10 Oct 07:20 & 23,200.00 & 23,200.00 & 5 & 0.0085 & 0.0087 & 0.014 & 0.0026 \\
NSW1 & 12 Oct 14:10 & 1,341.98 & 23,107.00 & 55 & 0.094 & 0.097 & 0.27 & 0.026 \\
QLD1 & 9 Oct 05:00 & 824.88 & 860.05 & 35 & 0.00074 & 0.00074 & 0.096 & 0.012 \\
SA1 & 9 Oct 07:15 & 1,001.00 & 3,210.98 & 10 & 0.033 & 0.033 & 0.0059 & 0.0018 \\
\bottomrule
\end{tabular}
}
\end{table}

\subsection{Confirmatory results}

\begin{figure}[!htbp]
  \centering
  \includegraphics[width=\textwidth,trim=0 0 0 27,clip]{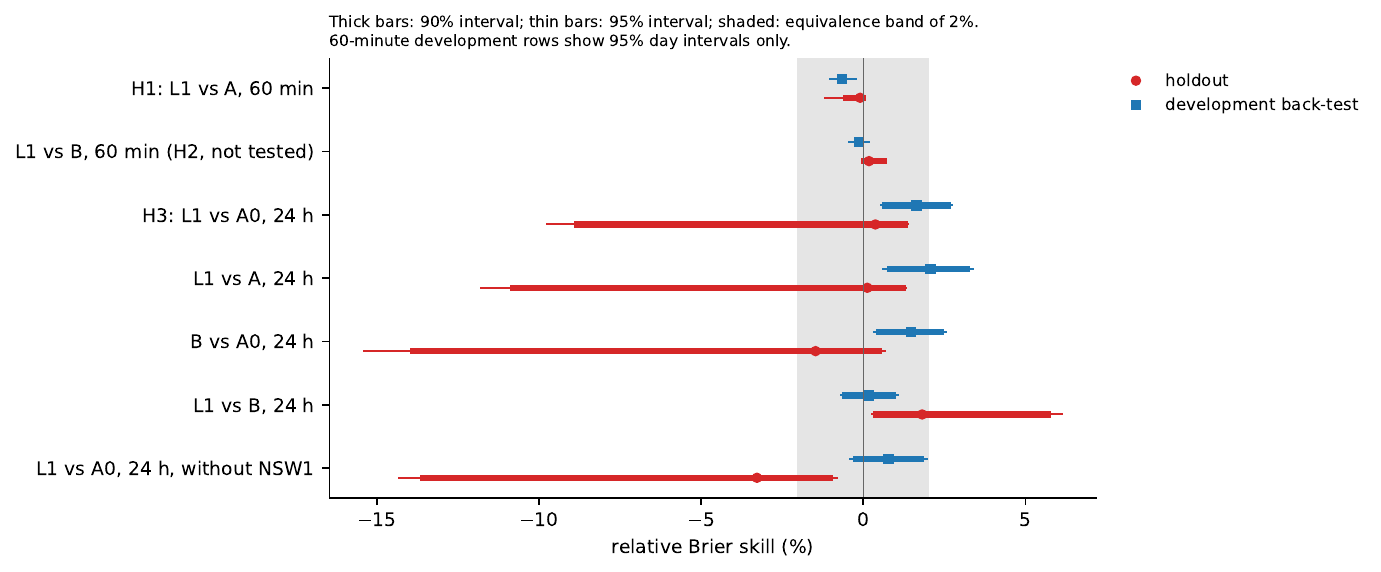}
  \caption{Relative Brier skill on the holdout (red) and in the development back-test (blue) for the three hypotheses and the other 24-hour comparisons listed with H3, except L2 against A0 (Table~\ref{tab:hold-24}). Thick bars are 90\% and thin bars 95\% intervals; the shaded area is the equivalence band. The 60-minute development rows show 95\% day intervals only.}
  \label{fig:hold-forest}
\end{figure}

The decisions were that H1 and H3 were not established and that H2 was not tested (Table~\ref{tab:hold-confirm} and Figure~\ref{fig:hold-forest}).

On the holdout, H1 was not established: the relative Brier skill of L1 against A was $-$0.09\% (90\% CI [$-$0.52, $-$0.02]; one-sided bootstrap $p = 0.9994$), inside the $\pm$2\% equivalence band. Because the final 60-minute text increments are single trees, H1 compares nearly identical forecasts; this does not show that text and numeric information are equivalent. The median absolute change in the logit relative to A was 0.0036 for L1, 0.0067 for B, 0.0073 for L2 and 0.0066 for A0, and it exceeded 0.05 at 1.1\%, 0.7\%, 0.3\% and 0.7\% of origins.

H2 was not tested, because H1 was not established and no level passed to H2.

H3 was not established: the relative Brier skill of L1 against A0 at 24 hours was $+$0.37\% (95\% CI [$-$9.74, 1.39]; one-sided bootstrap $p = 0.43$), an inconclusive result, which was the expected main outcome; it does not show that the notices are uninformative.

No test met its level, so no level was passed between hypotheses, and a Bonferroni reading at 0.025 gives the same decisions. The descriptive label of the untested comparison of L1 with B at 60 minutes, superior but below the smallest effect of interest (0.18\%, 90\% CI [0.01, 0.64]; bootstrap $p < 10^{-4}$ with 10,000 resamples), carries no confirmatory weight.

These intervals describe resampling of very few informative units and should not be read as sampling intervals for a population effect (post hoc reading). For H3 the distribution depends mostly on how often the two blocks with positives are drawn, and some resamples contain neither. For H1 and H2 the near-constant difference between nearly identical forecasts makes extreme p-values uninformative; the identical 90\% and 95\% limits of the comparison of L1 with B show the same discreteness. The decisions are unchanged.

\begin{table}[!htbp]
  \centering\footnotesize
  \caption{Confirmatory holdout results, five regions. Skill is the relative Brier skill of the first set against the second (positive favours the first), with percentile bootstrap intervals; $p$ is the one-sided bootstrap p-value and Level the initial one-sided level under the graphical Holm procedure (H2 receives a level only after H1 is established). The descriptive label does not replace the decision.}
  \label{tab:hold-confirm}
  \resizebox{\textwidth}{!}{\begin{tabular}{llllrllll>{\raggedright\arraybackslash}p{2.6cm}l}
\toprule
 & Comparison & Horizon & Bootstrap unit & Skill (\%) & 90\% CI & 95\% CI & $p$ & Level & Descriptive label & Decision \\
\midrule
H1 & L1 against A & 60 min & day clusters & $-$0.09 & [$-$0.52, $-$0.02] & [$-$1.18, $-$0.01] & 0.9994 & 0.025 & equivalent but slightly worse & not established \\
H2 & L1 against B & 60 min & day clusters & 0.18 & [0.01, 0.64] & [0.01, 0.64] & $<10^{-4}$ & 0 & superior but below the smallest effect of interest & not tested \\
H3 & L1 against A0 & 24 h & three-day blocks & 0.37 & [$-$8.81, 1.30] & [$-$9.74, 1.39] & 0.4335 & 0.025 & inconclusive & not established \\
\bottomrule
\end{tabular}
}
\end{table}

\subsection{Descriptive analyses fixed in the frozen protocol}

\begin{figure}[!htbp]
  \centering
  \includegraphics[width=0.9\textwidth]{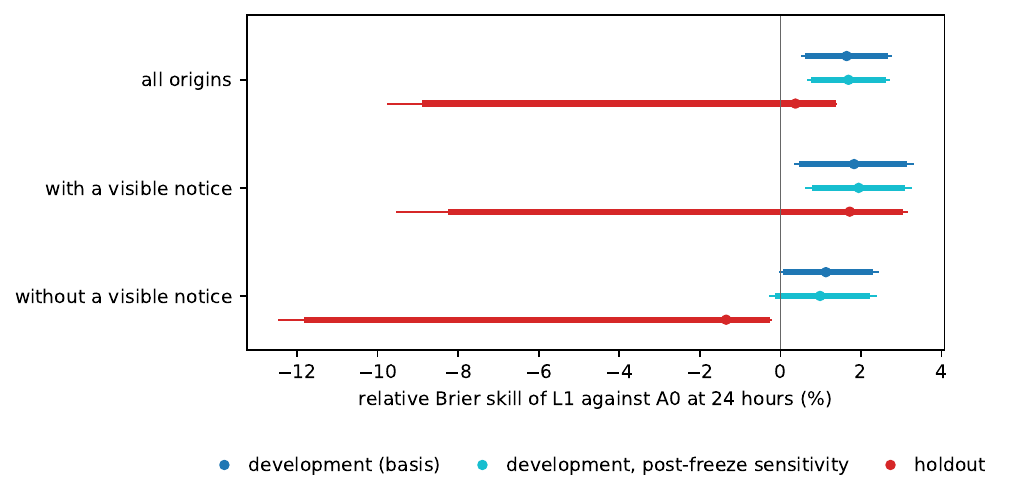}
  \caption{Descriptive split, fixed in the frozen protocol, of the relative Brier skill of L1 against A0 at 24 hours by the presence of a visible notice, on the holdout, in the development back-test and in its post-freeze refit (thick bars 90\%, thin bars 95\% three-day block intervals).}
  \label{fig:hold-strata}
\end{figure}

The frozen protocol attached to H3 a split by whether a notice was visible at the origin, a Murphy decomposition on fixed bins and two wording rules. At origins with a visible notice (47.9\% of the holdout, against 63.5\% in development) the relative skill of L1 against A0 at 24 hours was $+$1.72\% (90\% CI [$-$8.18, 2.98]; 95\% CI [$-$9.52, 3.13]; $p = 0.24$). At origins without one it was $-$1.35\% (95\% CI [$-$12.45, $-$0.23]), labelled inferior (Table~\ref{tab:hold-strata} and Figure~\ref{fig:hold-strata}). The rule that attributes a reduction to origins without notices therefore did not apply.

For the non-significant point-estimate reduction in the 24-hour Brier score of L1 relative to A0, the decomposition rule of the frozen protocol applies: the improvement comes mainly from recalibration and cannot be attributed to notice content.

\paragraph{Post hoc note.} In units of $10^{-5}$ (Table~\ref{tab:hold-murphy}), the reliability term fell by 5.38 and the resolution term rose by 5.18, a margin of 0.20, more than an order of magnitude below the fixed-bin residuals (L1 8.0, A0 3.9). With 20 equal-frequency bins the rule does not apply, although these bins fit worse (residuals 36 and 41); with the extended decomposition of \citet{stephensonTwoExtraComponents2008} it applies with a margin of 4.3. On this holdout the decomposition cannot separate recalibration from resolution. In the development back-test, with many more events, the resolution term of L1 exceeded that of A0 by about 180 while its reliability term was larger by 46.

\begin{table}[!htbp]
  \centering\footnotesize
  \caption{Descriptive split, fixed in the frozen protocol, of the 24-hour comparison of L1 with A0 by the presence of a visible notice at the origin: relative Brier skill (\%) with three-day block bootstrap intervals and one-sided $p$. The post-freeze refit is the development sensitivity analysis that ended every training set 24 hours before its fold.}
  \label{tab:hold-strata}
  \resizebox{\textwidth}{!}{\begin{tabular}{llrrrlll}
\toprule
Sample & Origins & $n$ & Positives & Skill (\%) & 90\% CI & 95\% CI & $p$ \\
\midrule
Holdout & All origins & 30,720 & 480 & 0.37 & [$-$8.81, 1.30] & [$-$9.74, 1.39] & 0.4335 \\
 & With a visible notice & 14,700 & 253 & 1.72 & [$-$8.18, 2.98] & [$-$9.52, 3.13] & 0.2417 \\
 & Without a visible notice & 16,020 & 227 & $-$1.35 & [$-$11.75, $-$0.33] & [$-$12.45, $-$0.23] & 0.9979 \\
\midrule
Development back-test & All origins & 365,313 & 51,198 & 1.65 & [0.68, 2.60] & [0.54, 2.74] & 0.0010 \\
 & With a visible notice & 231,937 & 38,673 & 1.83 & [0.53, 3.07] & [0.36, 3.29] & 0.0070 \\
 & Without a visible notice & 133,376 & 12,525 & 1.13 & [0.13, 2.22] & [$-$0.02, 2.43] & 0.0265 \\
\midrule
Post-freeze refit & All origins & 365,313 & 51,198 & 1.69 & [0.84, 2.55] & [0.69, 2.69] & 0.0005 \\
 & With a visible notice & 231,937 & 38,673 & 1.94 & [0.86, 3.02] & [0.63, 3.25] & 0.0010 \\
 & Without a visible notice & 133,376 & 12,525 & 0.99 & [$-$0.06, 2.14] & [$-$0.25, 2.38] & 0.0595 \\
\bottomrule
\end{tabular}
}
\end{table}

\begin{table}[!htbp]
  \centering\footnotesize
  \caption{Murphy decomposition of the 24-hour Brier score on the fixed reliability bins: Brier score $=$ REL $-$ RES $+$ UNC $+$ residual, where UNC is common to the models of a sample. REL is reliability (lower is favourable) and RES resolution (higher is favourable). The post-freeze refit is defined in Table~\ref{tab:hold-strata}.}
  \label{tab:hold-murphy}
  \begin{tabular}{llrrrrrr}
\toprule
Sample & Model & Brier $\times 10^{3}$ & REL $\times 10^{5}$ & RES $\times 10^{5}$ & Residual $\times 10^{5}$ & AUC & PR-AUC \\
\midrule
Holdout & L1 & 17.062 & 189.0 & 28.9 & 8.0 & 0.772 & 0.053 \\
 & A0 & 17.126 & 194.3 & 23.7 & 3.9 & 0.764 & 0.037 \\
 & A & 17.083 & 188.8 & 23.9 & 5.4 & 0.765 & 0.037 \\
\midrule
Development back-test & L1 & 102.044 & 82.0 & 1766.6 & $-$161.7 & 0.780 & 0.417 \\
 & A0 & 103.752 & 35.9 & 1582.3 & $-$129.0 & 0.766 & 0.390 \\
 & A & 104.205 & 19.1 & 1506.3 & $-$142.9 & 0.762 & 0.380 \\
\midrule
Post-freeze refit & L1 & 102.142 & 63.2 & 1766.6 & $-$133.1 & 0.779 & 0.414 \\
 & A0 & 103.899 & 39.2 & 1560.9 & $-$139.1 & 0.765 & 0.387 \\
 & A & 104.265 & 17.0 & 1512.4 & $-$128.8 & 0.763 & 0.378 \\
\bottomrule
\end{tabular}

\end{table}

\subsection{Listed comparisons and AEMO references}

At 60 minutes the comparisons listed with H1 lay inside the equivalence band (L1 against A0 $+$0.14\%, B against A0 $-$0.04\%, A0 against A $-$0.24\%). At 24 hours, where the L1 increment moved the logit relative to A0 by a median of 0.069 and by more than 0.05 at 88\% of origins, the listed comparisons were inconclusive except two that rest on five onsets or fewer (Table~\ref{tab:hold-24}). The exploratory comparison of L1 against B was $+$1.82\% (90\% CI [0.39, 5.68]), labelled superior; it is approximately the difference between L1 against A0 ($+$0.37\%) and B against A0 ($-$1.47\%). Without NSW1 (two onsets), L1 against A0 was $-$3.28\% (95\% CI [$-$14.31, $-$0.80]), labelled inferior.

\begin{table}[!htbp]
  \centering\footnotesize
  \caption{The 24-hour comparisons listed with H3 on the holdout: relative Brier skill with 90\% and 95\% three-day block bootstrap intervals and descriptive labels. The comparisons other than H3 are exploratory and rest on five onsets (two without NSW1).}
  \label{tab:hold-24}
  \begin{tabular}{>{\raggedright\arraybackslash}p{4.2cm}rll>{\raggedright\arraybackslash}p{4.4cm}}
\toprule
Comparison & Skill (\%) & 90\% CI & 95\% CI & Label \\
\midrule
L1 against A0 & 0.37 & [$-$8.81, 1.30] & [$-$9.74, 1.39] & inconclusive \\
L1 against A & 0.13 & [$-$10.79, 1.21] & [$-$11.78, 1.32] & inconclusive \\
B against A0 & $-$1.47 & [$-$13.90, 0.47] & [$-$15.39, 0.68] & inconclusive \\
L1 against B & 1.82 & [0.39, 5.68] & [0.26, 6.12] & superior \\
L2 against A0 & $-$0.65 & [$-$12.60, 0.56] & [$-$14.13, 0.69] & inconclusive \\
L1 against A0, without NSW1 & $-$3.28 & [$-$13.57, $-$1.01] & [$-$14.31, $-$0.80] & inferior \\
\bottomrule
\end{tabular}

\end{table}

Against AEMO's forecasts converted to probabilities, the comparisons of A were inconclusive at 60 minutes ($+$7.82\% over P5MIN, $-$2.09\% against PD) and labelled superior at 24 hours ($+$20.47\%, 90\% CI [5.35, 68.57]; $+$14.93\%, 90\% CI [2.15, 62.40]; Table~\ref{tab:hold-aemo}). The 24-hour references are AEMO's next-hour forecasts mapped to probabilities on the 24-hour training labels, whose base rate was 0.120. They carry no AEMO information beyond the next hour, and their mean forecast probability was 5.3 and 5.2 times the holdout base rate, against 2.6 for A (post hoc). These comparisons do not show that the numeric model makes more use of AEMO's information than AEMO's own forecasts.

\begin{table}[!htbp]
  \centering\footnotesize
  \caption{The numeric model A against AEMO's forecasts converted to probabilities and against the training-period climatology on the holdout: relative Brier skill (\%), percentile bootstrap intervals and descriptive labels. The 24-hour references are AEMO's next-hour forecasts mapped to probabilities and carry no AEMO information beyond the next hour; the label against climatology shows only that the climatology was a poor reference for the season.}
  \label{tab:hold-aemo}
  \resizebox{\textwidth}{!}{\begin{tabular}{>{\raggedright\arraybackslash}p{2.9cm}rll>{\raggedright\arraybackslash}p{2.0cm}rll>{\raggedright\arraybackslash}p{2.0cm}}
\toprule
 & \multicolumn{4}{c}{60 minutes (day clusters)} & \multicolumn{4}{c}{24 hours (three-day blocks)} \\
\cmidrule(lr){2-5}\cmidrule(lr){6-9}
Comparison & Skill & 90\% CI & 95\% CI & Label & Skill & 90\% CI & 95\% CI & Label \\
\midrule
A against P5MIN & 7.82 & [$-$2.42, 49.70] & [$-$4.44, 75.94] & inconclusive & 20.47 & [5.35, 68.57] & [3.53, 70.48] & superior \\
A against PD & $-$2.09 & [$-$10.86, 7.14] & [$-$18.20, 14.07] & inconclusive & 14.93 & [2.15, 62.40] & [0.69, 64.59] & superior \\
A against climatology & 16.15 & [4.89, 60.42] & [3.43, 86.22] & superior & -- & -- & -- & -- \\
A against P5MIN, without NSW1 & 21.52 & [8.85, 92.15] & [7.56, 93.80] & superior & 30.46 & [11.05, 66.18] & [8.85, 67.79] & superior \\
\bottomrule
\end{tabular}
}
\end{table}

\subsection{Secondary comparisons with jointly fitted models}

The frozen protocol names a secondary 60-minute design in which text actually enters the model. Numeric and text features are fitted together on the text window of October 2021 to July 2024, and the result is compared with the numeric model refitted on that window. These comparisons are outside the confirmatory family and received no level. They rest on 20 positive origins on three days, and their intervals are wide. Joint L1 against the refitted numeric model was $+$2.07\% (90\% CI [0.21, 10.36]; 95\% CI [0.01, 18.35]; $p = 0.0248$; sign-randomisation $p = 0.0567$), labelled superior, and joint L2 against joint L1 was $+$3.77\% (90\% CI [0.24, 15.87]; 95\% CI [0.05, 27.22]), also labelled superior. Joint B against the refitted numeric model was $+$1.32\% (90\% CI [$-$0.94, 5.11]) and joint L1 against joint B was $+$0.76\% (90\% CI [$-$3.14, 9.97]), both inconclusive. Without NSW1, which left 8 positive origins, all four comparisons were inconclusive. In the development back-test three of these comparisons were labelled equivalent: joint L1 against the refitted numeric model $-$0.51\% [$-$1.32, 0.34], joint B against it $-$0.02\% [$-$0.53, 0.51] and joint L2 against joint L1 $-$0.21\% [$-$0.55, 0.13]. Post hoc, the refitted numeric model had a higher Brier score than A (7.63 against 7.16, in units of $10^{-4}$; mean forecast probability 3.0 and 1.7 times the base rate), and joint L1 (7.47) and joint L2 (7.19) also stayed above A. A was fitted up to July 2026 and the refitted model up to July 2024, so this difference is not a test of the text, but the superior labels do not show a lower Brier score than the main numeric model. These labels do not change the decisions on H1 to H3.

\subsection{Sensitivity analyses}

The sensitivity analyses in the frozen protocol changed no decision (Table~\ref{tab:hold-sens}). For H1, a single seed, the literal administered-pricing rule, the 60-minute visibility delay and the exclusion of the first week gave $-$0.15\%, $-$0.09\%, $-$0.24\% and $-$0.09\%, all labelled equivalent but slightly worse. Only isotonic recalibration changed the label, to superior ($+$0.55\%, 90\% CI [0.01, 3.13]; 95\% CI [$-$0.01, 8.06]). For H3, a single seed gave $+$0.34\% (95\% CI [$-$9.84, 1.47]) and a day-cluster bootstrap $+$0.37\% ([$-$3.78, 2.27]), both inconclusive; the analysis without NSW1 is reported above. There is no visibility-delay sensitivity analysis for H3. The restriction to 15 notice types was fixed after the type counts of the captured holdout notices had been seen; without it, B against A was unchanged at $-$0.27\%.

\begin{table}[!htbp]
  \centering\scriptsize
  \caption{Sensitivity analyses in the frozen protocol, holdout (not confirmatory): relative Brier skill (\%), percentile bootstrap intervals (day clusters at 60 minutes and three-day blocks at 24 hours unless stated), one-sided $p$ and descriptive labels. In the notice-type analysis the two rows differ only beyond the second decimal.}
  \label{tab:hold-sens}
  \setlength{\tabcolsep}{4pt}
  \begin{tabular}{>{\raggedright\arraybackslash}p{3.0cm}>{\raggedright\arraybackslash}p{2.3cm}rlll>{\raggedright\arraybackslash}p{3.0cm}}
\toprule
Analysis & Comparison & Skill (\%) & 90\% CI & 95\% CI & $p$ & Label \\
\midrule
\multicolumn{7}{l}{\textit{60 minutes (day-cluster bootstrap)}} \\
Main analysis & L1 against A & $-$0.09 & [$-$0.52, $-$0.02] & [$-$1.18, $-$0.01] & 0.9994 & equivalent but slightly worse \\
 & L1 against B & 0.18 & [0.01, 0.64] & [0.01, 0.64] & $<10^{-4}$ & superior but below the smallest effect of interest \\
 & L1 against A0 & 0.14 & [0.03, 0.56] & [0.03, 1.11] & $<10^{-4}$ & superior but below the smallest effect of interest \\
\addlinespace
Single seed & L1 against A & $-$0.15 & [$-$0.74, $-$0.03] & [$-$1.53, $-$0.02] & 1.0000 & equivalent but slightly worse \\
 & L1 against B & 0.13 & [$-$0.003, 0.523] & [$-$0.01, 0.56] & 0.1118 & equivalent \\
 & L1 against A0 & 0.06 & [0.01, 0.29] & [0.01, 0.71] & 0.0013 & superior but below the smallest effect of interest \\
\addlinespace
Isotonic recalibration & L1 against A & 0.55 & [0.01, 3.13] & [$-$0.01, 8.06] & 0.0382 & superior \\
 & L1 against B & 0.49 & [0.003, 2.268] & [$-$0.01, 4.01] & 0.0448 & superior \\
 & L1 against A0 & 0.46 & [$-$0.03, 2.75] & [$-$0.06, 7.69] & 0.0743 & inconclusive \\
\addlinespace
Literal administered-pricing rule & L1 against A & $-$0.09 & [$-$0.52, $-$0.02] & [$-$1.20, $-$0.01] & 0.9993 & equivalent but slightly worse \\
 & L1 against B & 0.18 & [0.01, 0.63] & [0.01, 0.64] & $<10^{-4}$ & superior but below the smallest effect of interest \\
 & L1 against A0 & 0.14 & [0.03, 0.56] & [0.02, 1.08] & $<10^{-4}$ & superior but below the smallest effect of interest \\
\addlinespace
60-minute visibility delay & L1 against A & $-$0.24 & [$-$0.93, $-$0.02] & [$-$1.22, $-$0.01] & 1.0000 & equivalent but slightly worse \\
 & L1 against B & 0.09 & [0.01, 0.32] & [0.004, 0.469] & 0.0078 & superior but below the smallest effect of interest \\
 & L1 against A0 & 0.00 & [$-$0.11, 0.35] & [$-$0.14, 1.07] & 0.4590 & equivalent \\
\addlinespace
First week excluded & L1 against A & $-$0.09 & [$-$0.49, $-$0.02] & [$-$1.18, $-$0.01] & 0.9996 & equivalent but slightly worse \\
 & L1 against B & 0.18 & [0.01, 0.63] & [0.01, 0.64] & $<10^{-4}$ & superior but below the smallest effect of interest \\
 & L1 against A0 & 0.14 & [0.03, 0.55] & [0.03, 1.09] & $<10^{-4}$ & superior but below the smallest effect of interest \\
\addlinespace
Notice-type rule without restriction & B (all notice types) against A & $-$0.27 & [$-$1.15, $-$0.03] & [$-$1.74, $-$0.02] & 1.0000 & equivalent but slightly worse \\
 & B (fifteen retained types) against A & $-$0.27 & [$-$1.15, $-$0.03] & [$-$1.74, $-$0.02] & 1.0000 & equivalent but slightly worse \\
\midrule
\multicolumn{7}{l}{\textit{24 hours (three-day block bootstrap unless stated)}} \\
Single seed & L1 against A0 & 0.34 & [$-$8.93, 1.37] & [$-$9.84, 1.47] & 0.4323 & inconclusive \\
\addlinespace
Day-cluster bootstrap & L1 against A0 & 0.37 & [$-$2.53, 1.95] & [$-$3.78, 2.27] & 0.3983 & inconclusive \\
\addlinespace
Without NSW1 & L1 against A0 & $-$3.28 & [$-$13.57, $-$1.01] & [$-$14.31, $-$0.80] & 1.0000 & inferior \\
\addlinespace
Bonferroni reading at 0.025 & L1 against A0 & -- & -- & -- & -- & not established \\
\bottomrule
\end{tabular}

\end{table}

\subsection{Event-level description}

With the frozen alarm thresholds, A and L1 each detected 1 of the 5 onsets with 6 false alarms, P5MIN 2 with 23 and PD 0 with 12 (Table~\ref{tab:hold-alarms}). Lead times rest on at most two detections per model and are not compared.

\begin{figure}[!htbp]
  \centering
  \includegraphics[width=\textwidth,trim=0 0 0 30,clip]{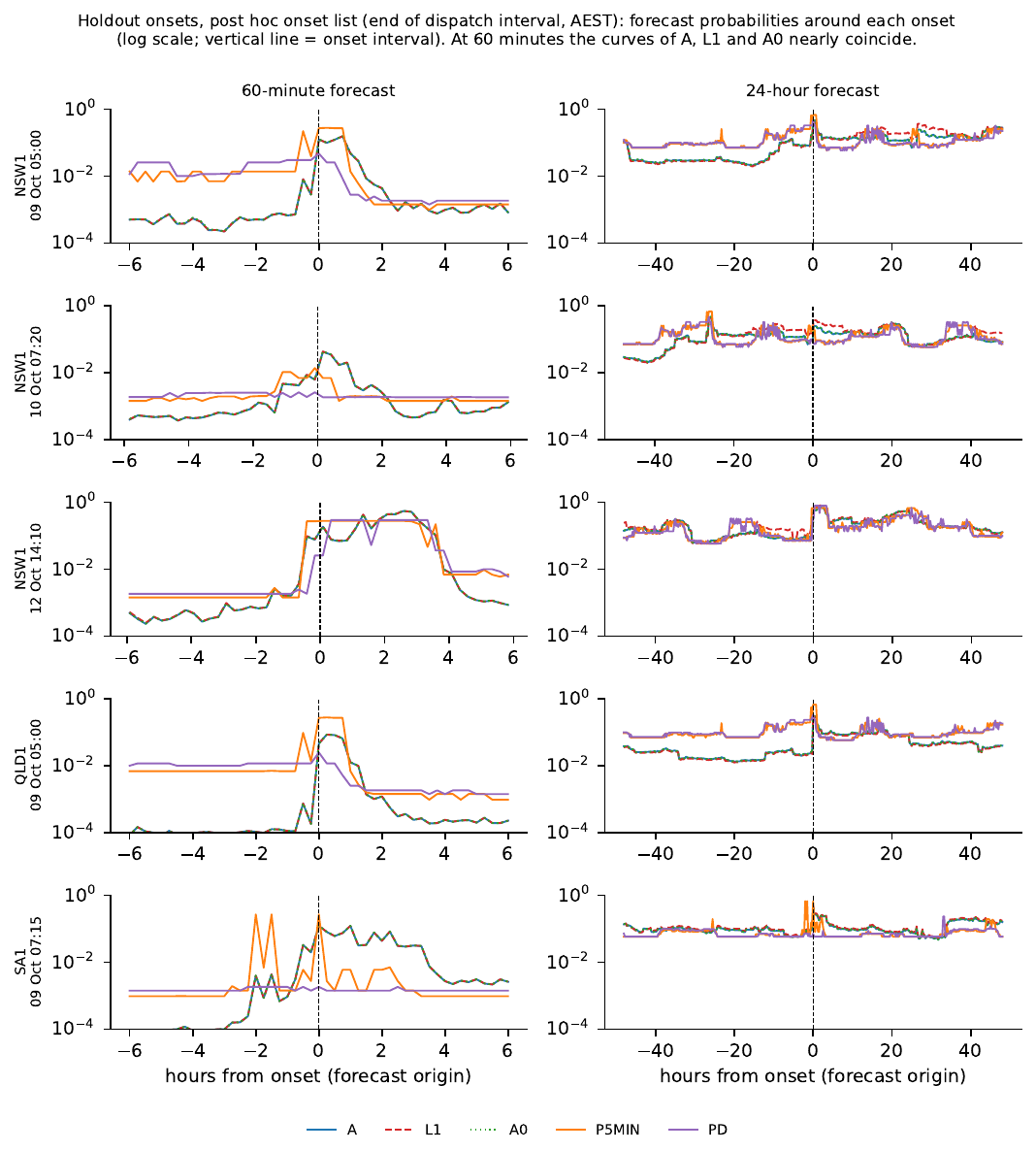}
  \caption{Post hoc description: forecast probabilities around the five holdout onsets (end of the dispatch interval, AEST), at 60 minutes (left) and 24 hours (right), on a log scale; the vertical line marks the onset interval. At 60 minutes the curves of A, L1 and A0 nearly coincide.}
  \label{fig:hold-events}
\end{figure}

The secondary outcomes were fitted with a single seed and carry no intervals. They rest on five events and cannot be interpreted. The 50\% and 90\% pinball losses of the hourly peak price were 11.33 and 10.54 for A, 11.35 and 10.82 for B, and 11.31 and 10.83 for L1. The mean absolute error of the minutes above 600~AUD/MWh within the hour, which is not the episode duration, was 0.0280, 0.0288 and 0.0280 minutes, and the mean squared error 0.679, 0.626 and 0.675 minutes squared. Post hoc, predicting no such minutes gave a lower mean absolute error (0.0234 minutes) and a higher mean squared error (0.791 minutes squared) than all three. Figure~\ref{fig:hold-events} shows the forecasts around each onset, a post hoc description.

\begin{table}[!htbp]
  \centering\footnotesize
  \caption{Event-level alarms on the holdout with the alarm thresholds frozen before opening, five regions summed. Lead times rest on at most two detections per model and are not compared.}
  \label{tab:hold-alarms}
  \begin{tabular}{lrrrrr}
\toprule
Model & Onsets & Detected & Missed & False alarms & Lead time of detections (min) \\
\midrule
A & 5 & 1 & 4 & 6 & 30 \\
B & 5 & 1 & 4 & 6 & 30 \\
L1 & 5 & 1 & 4 & 6 & 30 \\
L2 & 5 & 1 & 4 & 6 & 30 \\
P5MIN & 5 & 2 & 3 & 23 & 30 \\
PD & 5 & 0 & 5 & 12 & -- \\
\bottomrule
\end{tabular}

\end{table}

\section{Post hoc descriptions}\label{sec:posthoc}

Everything in this section was specified and computed after the holdout was opened. It is descriptive, adds no hypothesis or label and changes no decision.

\subsection{The holdout}

\begin{figure}[!htbp]
  \centering
  \includegraphics[width=\textwidth,trim=0 0 0 30,clip]{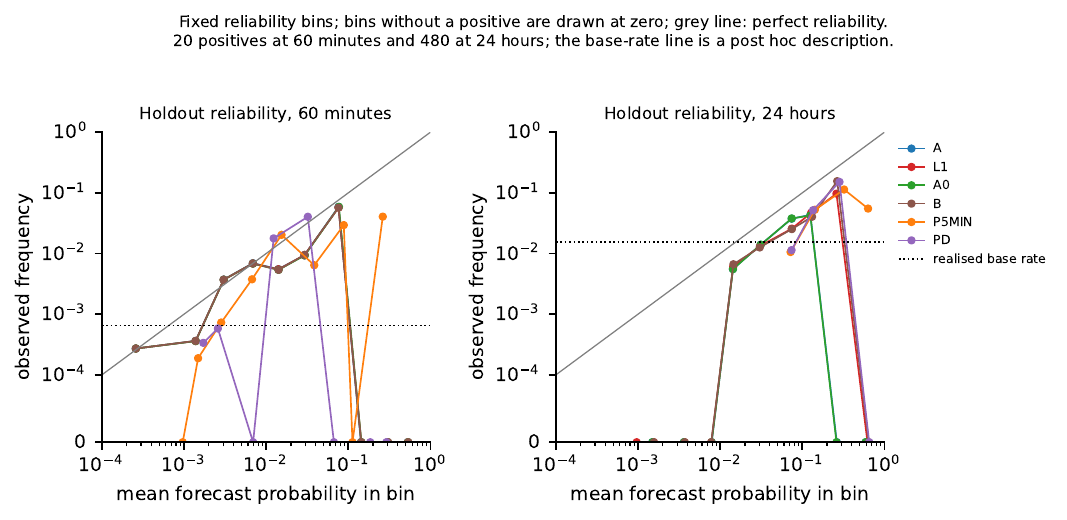}
  \caption{Holdout reliability diagrams at 60 minutes and 24 hours on the fixed reliability bins; bins without a positive are drawn at zero, and the grey line marks perfect reliability. The holdout had 20 positive origins at 60 minutes and 480 at 24 hours. The line at the realised base rate is a post hoc description.}
  \label{fig:hold-reliability}
\end{figure}

\begin{figure}[!htbp]
  \centering
  \includegraphics[width=\textwidth,trim=0 0 0 18,clip]{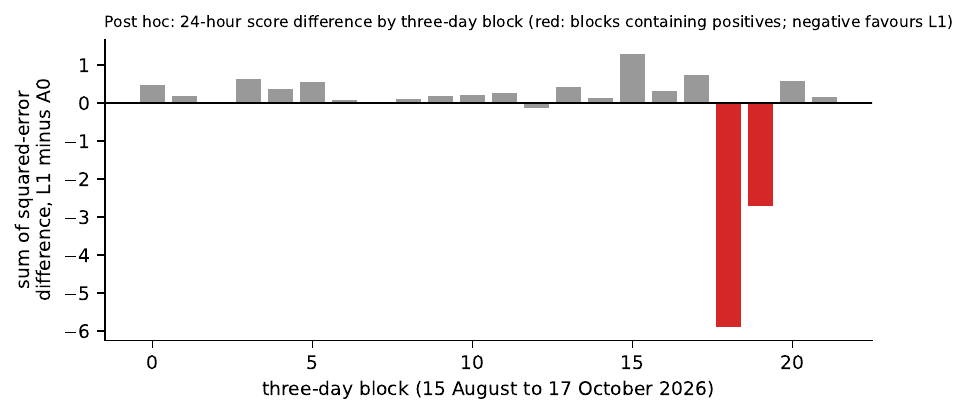}
  \caption{Post hoc description: sum of the squared-error difference between L1 and A0 at 24 hours in each three-day block of the holdout (red: blocks containing positives; negative values favour L1).}
  \label{fig:hold-blocks}
\end{figure}

The mean forecast probability of A was about 1.7 times the base rate at 60 minutes and 2.6 times at 24 hours (Figure~\ref{fig:hold-reliability}). Summed over three-day blocks, the squared-error difference between L1 and A0 at 24 hours was $-$8.60 in the two blocks with positives and $+$6.63 in the other 20 (Figure~\ref{fig:hold-blocks}).

\subsection{Timing of notices against numeric signals}

This description uses only development data: the 967 onsets of the nine back-test folds (July 2022 to July 2024, five regions) and the notices of October 2021 to July 2024. A relevant notice was of a retained type, assigned to the onset's region or the whole market, not a cancellation and visible under Equation~\eqref{eq:visible} with a delay of 25 minutes. A numeric alarm was the numeric model's probability at or above its frozen threshold, a P5MIN price above 600~AUD/MWh in its next hour, or a PD price above 300~AUD/MWh for the onset half-hour.

For 464 of 967 onsets (48.0\%) a relevant notice had become visible within six hours before the first origin covering the onset, against 36.5\% of matched times in the same month, region and hour with no onset within 24 hours (ratio 1.31 [1.19, 1.44]). Of the 398 onsets at which both a notice and a numeric alarm appeared in those six hours, an alarm came before or at the same origin as the notice for 84.7\% [79.3, 89.1]. The order depended on the signal (Table~\ref{tab:posthoc-order}). A PD price above 300~AUD/MWh came before or with the notice for 93.1\% of 334 onsets, but the numeric model's own alarm did so for only 29.8\% of 366; the notice came before that alarm for 70.2\% [63.8, 77.0], against 62.9\% expected had notices arrived as at the matched times. Because relevant notices were also more frequent before onsets, these differences reflect how often notices appeared as well as when they appeared; for 69.8\% of the 929 onsets compared, a numeric alarm was already present just before the six-hour window. These shares describe timing at these alarm levels. They do not show whether notices carried information that the numeric state lacked.

\begin{table}[!htbp]
  \centering\footnotesize
  \setlength{\tabcolsep}{3.5pt}
  \caption{Post hoc description: order of relevant notices and numeric signals in the six hours before spike onsets in the development period. For onsets at which both a notice and the signal appeared, the share at which the notice came first, the share expected had notices arrived as at the matched times, their difference in percentage points and the share at which the signal came before or at the same origin, with 95\% day-block bootstrap intervals. The intervals do not include the sampling variation of the matched times. The first row pools the signals; the other rows list each signal.}
  \label{tab:posthoc-order}
  \begin{tabular}{>{\raggedright\arraybackslash}p{2.9cm}rrlrrlrl}
\toprule
Numeric signal & \makecell{Onsets\\with both} & \makecell{Notice\\first (\%)} & 95\% CI & \makecell{Expected\\(\%)} & \makecell{Difference\\(pp)} & 95\% CI & \makecell{Signal first\\or same (\%)} & 95\% CI \\
\midrule
Any numeric alarm (pooled) & 398 & 15.3 & [10.9, 20.7] & 13.7 & 1.7 & [$-$1.0, 4.8] & 84.7 & [79.3, 89.1] \\
Numeric model's alarm & 366 & 70.2 & [63.8, 77.0] & 62.9 & 7.3 & [2.4, 12.0] & 29.8 & [23.0, 36.2] \\
P5MIN price above 600 AUD/MWh & 339 & 69.3 & [61.5, 77.5] & 62.0 & 7.3 & [2.4, 12.4] & 30.7 & [22.5, 38.5] \\
PD price above 300 AUD/MWh for the onset half-hour & 334 & 6.9 & [4.1, 10.5] & 6.0 & 0.9 & [$-$0.8, 2.9] & 93.1 & [89.5, 95.9] \\
P5MIN reference alarm at its frozen threshold & 332 & 67.5 & [58.6, 76.7] & 61.2 & 6.3 & [0.7, 11.9] & 32.5 & [23.3, 41.4] \\
PD reference alarm at its frozen threshold & 303 & 63.4 & [53.4, 74.3] & 57.2 & 6.2 & [0.7, 12.1] & 36.6 & [25.7, 46.6] \\
\bottomrule
\end{tabular}

\end{table}

\section{Discussion}\label{sec:discussion}

\subsection{What the evidence supports}

In the development back-test at 60 minutes, neither LLM-extracted nor rule-based notice features lowered the Brier score of a numeric model that includes AEMO's forecasts: L1 against A was $-$0.66\% [$-$1.04, $-$0.24] (development out-of-sample), inside the $\pm$2\% band. On the holdout H1 was not established ($-$0.09\%), but the final 60-minute text increments are single trees, so H1 compared nearly identical forecasts. This does not show that text and numeric information are equivalent. In the secondary jointly fitted models, where text enters the model, joint L1 was labelled superior to the numeric model refitted on the same window on the holdout and equivalent to it in development (Section~\ref{sec:holdout}); the holdout value rests on 20 positive origins, joint B was inconclusive, and the refitted model had a higher Brier score than A (post hoc).

At 24 hours the question remains open. In development, L1 against the text-free control A0 was $+$1.65\% [0.54, 2.74], and the decomposition placed the reduction in resolution. The rule-based set B reached almost the same value ($+$1.47\%), which shows that the effect is not specific to the LLM, not that it is unrelated to the notices. Attribution to notice content is not established: L1 also had a positive point estimate at origins without a visible notice ($+$1.13\%, with an interval that includes zero), and every text increment shifted the forecasts when no notice was visible. A 24-hour gain from notice features may also reflect public numeric information that A omits (Section~\ref{sec:features}); this was not tested. On the holdout, H3 was not established ($+$0.37\%, 95\% CI [$-$9.74, 1.39], inconclusive). For this non-significant point estimate, the rule in the frozen protocol applies: the improvement comes mainly from recalibration and cannot be attributed to notice content. A post hoc check found that the holdout decomposition cannot separate recalibration from resolution. The holdout neither established nor excluded a 24-hour effect, and it does not show that notices are uninformative.

\subsection{Why notices may add little within the hour}

A possible explanation, which this design does not test, is that much of what notices report is already in the numeric state: many reserve conditions, outages, constraint changes and interventions also affect the P5MIN and PD forecasts. In a post hoc description of the development period, a numeric alarm came before or at the same origin as the notice at 84.7\% [79.3, 89.1] of the onsets where both appeared, but the share was 93.1\% for a PD price above 300~AUD/MWh and 29.8\% for the numeric model's own alarm (Table~\ref{tab:posthoc-order}). These shares describe timing at these alarm levels, not whether notices carried information that the numeric state lacked. L1 and B were equivalent in development. Given the high agreement between LLM and rules on most fields (Section~\ref{sec:features}), this may reflect shared extraction rather than the content of the notices; the design cannot separate the two.

\subsection{Relation to the studies reviewed here}

Several studies reviewed in Section~\ref{sec:related} report lower errors with text, in different settings; our results neither contradict them directly nor show that those reductions are spurious. Among those studies, only \citet{huangLLMEnhancedShortTermElectricity2025} tested the text increment, and over all intervals of its 22 test days the difference was not distinguishable from zero, while it was distinguishable in intervals above \$100/MWh, a subset defined by realised prices. \citet{luLargeLanguageModelBased2025} reported lower errors than AEMO's forecast in one test month, for a full system of which news sentiment was one part. Our results suggest comparing a reduction attributed to an LLM with simpler text processing and with a numeric model that includes the operator's forecasts.

\subsection{Reading the holdout (post hoc)}

Holdout base rates were far below development ones (Table~\ref{tab:base-rates}). The 24-hour reliability terms (189 to 194, in units of $10^{-5}$) far exceeded the resolution terms (24 to 29), while the AUC (0.76 to 0.77) stayed close to its development values (0.76 to 0.78; Table~\ref{tab:hold-murphy}). The scores therefore say more about calibration under a shifted base rate than about discrimination. A rare-event holdout should be sized by its expected number of target events, counted with the rule of the frozen labels, not by calendar time: a projection using a different count gave 20--34 onsets, against five observed.

\subsection{Practical implication}

A participant who wants an early warning of spikes within the hour should start from the operator's forecasts and the current numeric state. At longer horizons, AEMO's forward PD and PASA data should be compared with text before skill is attributed to text. This rests on development evidence: A, which takes P5MIN and PD as inputs, had pooled relative skill of $+$3.66\% over converted P5MIN and $+$6.29\% over converted PD. The P5MIN comparison was inconclusive on the five folds not used for hyperparameter selection, and both were inconclusive at 60 minutes on the holdout. Because base rates shift, a deployed forecaster needs its calibration monitored and its probabilities recalibrated. We assess forecast quality, not value to a decision maker.

\subsection{Limitations}

Appendix~\ref{app:deviations} lists the deviations from the frozen protocol; these disclosures bear most on how the evidence should be read. The protocol was frozen and hashed locally before opening and was not deposited in a public registry. The holdout is retrospective: its onsets had occurred before the first freeze, and that its market values were not read before the final freeze rests on our own records. Many descriptive holdout comparisons outside the confirmatory family, including the listed, sensitivity and secondary joint comparisons, were labelled without multiplicity control. H3 was added after development signals had been seen, many of the 44 amendments followed development results, and hyperparameters were selected partly on development folds. The notice-type restriction was fixed after holdout type counts had been seen, and earlier checks touched holdout metadata. We were not blind to the NSW1 notices of October 2021 to July 2024, which the first author's earlier work had used. The visibility delay rests on 43 measured notices and is a proxy for other periods, and H3 has no sensitivity analysis for it. The data used here contain no notice text from August 2024 to mid-August 2026, so text features learned up to July 2024 are applied from August 2026. At 24 hours A has no AEMO forecast beyond the next hour and no PASA data. The final 60-minute text increments are single trees. The L2 principal components were fitted on all training notices, including those after each development fold. There is no public evidence on the accuracy of the extracted fields, and we did not validate them against human labels. The 24-hour training labels overlap the first hours of each fold; a post-freeze refit gave $+$1.69\% [0.69, 2.69], a change classified as small by a rule fixed beforehand. With 5 onsets on 3 days and 24-hour positives in 2 of 22 three-day blocks, the frozen power estimate overstated the power of the holdout. We studied one open model of moderate size, one spike threshold and one market design, so the results may not transfer.

\section{Conclusion}\label{sec:conclusion}

We asked whether AEMO market notices, read once by a local LLM and used only after they became visible, improve short-horizon spike forecasts in five NEM regions. In the development back-test, which is development out-of-sample evidence, they did not lower the 60-minute Brier score of a numeric model that includes AEMO's forecasts. At 24 hours the reduction was shared by rule-based features, and its attribution to notice content was not established. On the retrospective single-use holdout, under a locally frozen protocol, H1 was not established ($-$0.09\%), H2 was not tested and H3 was not established ($+$0.37\%, 95\% CI [$-$9.74, 1.39], inconclusive). With 5 onsets on 3 days and 24-hour positives in 2 of 22 three-day blocks, the holdout neither established nor excluded a 24-hour effect. Whether notices add information at 24 hours remains open. A replication should set its length by the expected number of onsets.

\section*{Data and code availability}

The market data and market notices are public data of the Australian Energy Market Operator (AEMO), taken from the monthly archives of the Market Management System Data Model and from the NEMweb report directories. AEMO grants general permission to use its published material for any purpose, provided that the material and AEMO as its author are accurately attributed \citep{aemoCopyrightPermissions2026}. Market notices of the holdout period were obtained from the public NEMweb directory of current notices, by a one-off backfill and then live polling, both from shortly after midnight AEST on 13 October 2026. That directory keeps only about the most recent 60 days, so the holdout notices cannot be assumed to remain available there. The frozen protocol, whose text is mainly in Chinese, and the code and predictions are held by the first author and are available on reasonable request. They are not publicly deposited, so readers cannot verify the frozen content or its timing from this paper; the hashes in Table~\ref{tab:frozen-ids} allow files obtained on request to be checked.

\section*{Relation to other work}

This paper is adapted from Liang Lin's honours thesis at The University of Sydney (2026), which also reports earlier single-region studies that are not part of the evidence here. The results of this paper were obtained under the protocol described in Section~\ref{sec:evaluation}.

\section*{Use of generative AI}

Language models had two separate roles in this work. A local open model, a community text-only build of Qwen3.5-9B \citep{qwenTeamQwen35ModelCard2026}, is a research object: it extracted features from market notices, and the value of those features is what this paper evaluates. Separately, Anthropic's Claude and OpenAI's ChatGPT and Codex were used to assist with code development, grammatical corrections, and improving the clarity and conciseness of the manuscript. All AI-generated code and AI-assisted edits were reviewed and verified by the first author. All authors take full responsibility for the final content of this paper.
\section*{Author contributions}

Liang Lin designed the study with supervision, directed and checked the experiments, and prepared the paper using the tools described above. Jianguo Zhu supervised the research. Gang Lei, Allen Wang and Xin Lu contributed through discussions of the research.

\section*{Competing interests}

Allen Wang is employed by Wood Mackenzie. Jianguo Zhu, Gang Lei and Allen Wang are co-authors of \citet{huangLLMEnhancedShortTermElectricity2025}, and Xin Lu is a co-author of \citet{luLargeLanguageModelBased2025}; both works are discussed in Section~\ref{sec:related}.

\section*{Acknowledgement}

We thank AEMO for making its market data and market notices publicly available. The views expressed and any errors are our own.

\appendix
\section{Deviations and disclosures relative to the frozen protocol}\label{app:deviations}

The protocol was frozen locally in two versions (Section~\ref{sec:evaluation}) and was not deposited in a public registry. Table~\ref{tab:frozen-ids} gives the identifiers of the frozen material. These hashes show that files are unchanged. They do not show when the files were frozen. The holdout was opened once, on 20 October 2026, and scoring ended at 01:39:32 AEST. Analyses specified after the holdout was opened are labelled post hoc wherever they appear. No item below changes a confirmatory decision.

\begin{table}[!htbp]
  \centering\footnotesize
  \caption{Identifiers of the frozen material and of the confirmatory output. The prompt hash covers the prompt string followed by the token limits.}
  \label{tab:frozen-ids}
  \begin{tabular}{ll}
    \toprule
    Item & Identifier \\
    \midrule
    Manifest of file hashes (SHA-256) & {\scriptsize\texttt{20aff96243bd6aae1913683ffd7042f0eff3c2a3db15671e2eb6953df8d29ff9}} \\
    Prompt and token limits (SHA-256) & {\scriptsize\texttt{86d01917a40ab956be9bdaeb1caceafbccfa62e8be337bc0225b5853c1c14f2f}} \\
    Confirmatory output file (SHA-256) & {\scriptsize\texttt{15b81749fa3208bf8eadcfabae77fea8bb35f1a641527fd1273e9e033afb231e}} \\
    Text-only build of Qwen3.5-9B & \texttt{principled-intelligence/Qwen3.5-9B-text-only} \\
    Model revision & \texttt{2ba8a2e101fb96c67b3ce5a83beddc64c23fff43} \\
    \bottomrule
  \end{tabular}
\end{table}

\paragraph{What the two versions fixed.} The first version fixed the question, the feature sets A, B, L1 and L2, the forecaster and its candidates, the development back-test, H1 and H2, the scoring rule, the equivalence band, the descriptive labels and the holdout period. The second version added H3 and incorporated the amendments.

\paragraph{Amendments.} The 44 dated amendments between the two freezes fall into five groups:
\begin{itemize}
  \item data and the as-of rule: the source of holdout numeric data, the use of first-seen times to measure the delay, delays above the planned limit, notices captured during polling interruptions, the notice-type restriction and the completion of PD forecasts for the holdout;
  \item corrections of implementation errors found during development, among them rows not sorted by time, an inconsistency between the offset used in training and in prediction, and the handling of missing values;
  \item model and feature details left open by the first version: the number of seeds, the rule for cancelled lack-of-reserve notices, the removal of the numeric logit from the increment features and the number of embedding components;
  \item added or clarified analyses: the 24-hour exploratory analysis that later became H3, the text-free control A0, a placebo comparison in the development period only, secondary calibration measures, the disclosure and reading rule for near-zero 60-minute increments, and the split of the holdout by the presence of notices;
  \item the opening procedure: a two-step script that produces forecasts before reading any label, its failure paths, the rules for resuming a scoring run, the earliest time of opening and the single confirmatory output file.
\end{itemize}
Many amendments were written after development results had been seen. The most consequential, the addition of H3, was motivated by a development result. None was informed by holdout market values or the content of holdout notices; the visibility delay and the notice-type restriction used the timing and type counts of holdout notices. Table~\ref{tab:deviations} lists deviations and post-opening disclosures, and Section~\ref{sec:discussion} lists the limitations that bear most on how the evidence should be read.

\begingroup
\footnotesize
\setlength{\LTpre}{6pt}
\setlength{\LTpost}{6pt}
\begin{longtable}{>{\raggedright\arraybackslash}p{0.6cm}>{\raggedright\arraybackslash}p{10.0cm}>{\raggedright\arraybackslash}p{2.3cm}>{\raggedright\arraybackslash}p{1.3cm}}
\caption{Deviations and post-opening disclosures relative to the frozen protocol.}\label{tab:deviations}\\
\toprule
No. & Item & Type & Affects a decision? \\
\midrule
\endfirsthead
\toprule
No. & Item & Type & Affects a decision? \\
\midrule
\endhead
\bottomrule
\endfoot
1 & The frozen text contains two out-of-date details. It quotes the development 24-hour skill of L1 against B as $+$0.06\%, a value computed with the earlier 30-minute visibility delay; with the frozen 25-minute delay it is $+$0.18\%, and $+$0.25\% in the post-freeze refit. A section heading names an earlier version of the opening script than the one frozen and run. Neither was changed, because the text is frozen. & Out-of-date text & No \\
2 & The descriptive split by the presence of notices was computed by a separate script after scoring. Its first run stopped while writing its output file; after a one-line fix it was rerun, and all tables were identical. The step lines of the opening script, which contain no scores, were visible on the console during the run, although the procedure allowed only the last line to be read, and only after completion. & Procedural & No \\
3 & A post-freeze, pre-opening 24-hour refit, with training sets ending 24 hours before each fold, is listed beside H3 and in the development reference of the notice split, for description only. This was decided on 19 October 2026, before opening. H3 rests on the original back-test. & Added descriptive comparison & No \\
4 & The holdout had far fewer events than the design assumed: 5 onsets, against 20--34 in a design-stage projection from monthly onset counts of 2023--2025, counted on trading prices with a definition that differs from the frozen label (Section~\ref{sec:holdout}). & Disclosure & No \\
5 & The 25-minute visibility delay rests on 43 notices observed live. The excluded notices, the percentile for retained types and the unmeasured period are given in Section~\ref{sec:data}. & Limitation & No \\
6 & Notice extraction ran in a separate software environment from the numeric pipeline. The environment of the training-period extraction was not recorded at the time. The environment of the holdout extraction reproduced the stored training-period outputs byte for byte on four training notices before the freeze and on 40 before opening (raw output, parse and embedding), which indicates, but does not prove, that the two are equivalent. & Disclosure & No \\
7 & The 24-hour P5MIN and PD references are AEMO's next-hour forecasts, mapped to probabilities on the 24-hour training labels, a construction frozen with the models (Section~\ref{sec:holdout}). They are not an AEMO 24-hour forecast. & Disclosure & No \\
8 & The secondary outcomes were reported as point values from a single seed, without the intervals named in the frozen protocol (Section~\ref{sec:holdout}). & Reporting & No \\
9 & Two exploratory development results made before the final freeze are reported although the frozen protocol does not list them as outputs: the agreement of the LLM with the fixed rules (Section~\ref{sec:features}) and the run without AEMO's forecasts (Section~\ref{sec:discussion}). & Added description & No \\
\end{longtable}
\endgroup

\bibliographystyle{IEEEtranN}
\bibliography{references}
\end{document}